\documentclass[reprint,superscriptaddress, amsmath,amssymb,aip,pof,onecolumn,nofootinbib]{revtex4-2}
\usepackage{xr,mathtools}
\usepackage{hyperref}
\usepackage[dvipsnames]{xcolor}
\usepackage[T1]{fontenc}
\usepackage{lmodern}
\usepackage{textcomp}
\usepackage{dcolumn}
\usepackage{tabularx}
\usepackage[framemethod=tikz]{mdframed} 
\usepackage{listings,xcolor}
\usepackage{empheq}

\DeclareMathAlphabet{\mathbi}{OML}{cmm}{b}{it} 
\newcommand{\bu}{\mbox{\boldmath$u$}}
\newcommand{\bom}{\mbox{\boldmath$\omega$}}
\newcommand{\bpsi}{\mbox{\boldmath$\psi$}}

\newcolumntype{L}{>{\hsize=2.3\hsize}>{$}X<{$}} 
\newcolumntype{N}{>{\hsize=1\hsize}X} 
\begin{document}

\title{Early-time resonances in the three-dimensional wall-bounded axisymmetric Euler 
and related equations}
\author{Sai Swetha Venkata Kolluru}
\author{Rahul Pandit}
\affiliation{Department of Physics, Indian Institute 
 of Science, Bengaluru, India - 560012}

\begin{abstract}
        We investigate the complex-time analytic structure of solutions of the 3D-axisymmetric, wall-bounded, incompressible Euler equations, by starting with the initial data proposed in Ref.~\onlinecite{luo2014potentially}, to study a possible finite-time singularity. We use our pseudospectral Fourier-Chebyshev method~\cite{kolluru2022insights}, with quadruple-precision arithmetic, to compute the time-Taylor series coefficients of the flow fields, up to a high order. We show that the resulting approximations display \textit{early-time resonances}; the initial spatial location of these structures is different from that for the \textit{tygers}, which we have obtained in Ref.~\onlinecite{kolluru2022insights}. We then perform asymptotic analysis of the Taylor-series coefficients, by using generalised ratio methods, to extract the location and nature of the convergence-limiting singularities and demonstrate that these singularities are distributed around the origin, in the complex-$t^2$ plane, along two curves that resemble the shape of an eye. We obtain similar results for the 1D wall-approximation (of the full 3D-axisymmetric Euler equation) called the 1D HL model, for which we use Fourier-pseudospectral methods to compute the time-Taylor series coefficients of the flow fields. Our work examines the link between tygers, in Galerkin-truncated pseudospectral studies, and early-time resonances, in truncated time-Taylor expansions of solutions of PDEs, such as those we consider. 

\end{abstract}

\maketitle

\section{Introduction}

The regularity problem for hydrodynamical partial differential equations (PDEs), such as the three-dimensional (3D) Euler and the 3D Navier-Stokes equations~\cite{constantin1988navier,doering1995applied,constantin2005euler}, has been the subject of several theoretical and numerical investigations over the last few decades. However, it is still not known whether the solutions of these PDEs, starting from smooth initial data, retain smoothness, for all time, or whether they lose regularity, at a finite time, leading to a \textit{finite-time singularity} (often called \textit{blowup} in common parlance)~\cite{gibbon2008euler}. The breakdown of the regularity of solutions of the  3D Euler equations has been conjectured to play a key role in  understanding of the structure of Navier--Stokes turbulence [see, e.g., Refs.~\onlinecite{onsager1949statistical,constantin1994onsager,eyink2006onsager,eyink2008dissipative,de2014dissipative}]. 

The singularity problem for hydrodynamical PDEs has often been investigated by examining the analytic structure of solutions in the \textit{complex-time domain}. The solution of the governing PDE is represented as a Taylor series in integer powers $n$ of the time $t$. The asymptotic behavior of these Taylor coefficients is determined by the dominant complex-time singularities that lie closest to the point about which the Taylor expansion is made. 
Early investigations of the finite-time-singularity problem for the 3D Euler equation, starting from Taylor--Green initial data~\cite{morf1980spontaneous}, used methods of asymptotic analysis that were developed originally for the study of singularities in critical phenomena~\cite{gaunt1974asymptotic}, e.g., the ratio test~\cite{domb1957susceptibility} and Pad{\'e}-approximant methods~\cite{bender2013advanced}. These methods were used in conjunction with high-precision arithmetic to study the analytic structure of Euler flows~\cite{morf1980spontaneous,brachet1983small,brachet1984taylor,pelz1997evidence}, high-Reynolds-number turbulent flows~\cite{morf1981analytic}, and related hydrodynamical problems~\cite{frisch1981intermittency,meiron1982analytic,ohkitani2000numerical}. 

Over the years, direct-numerical-simulation (DNS) 
searches for possible singular solutions of hydrodynamical PDEs have grown apace with advances in high-performance computing and state-of-the-art numerical schemes have been developed to capture the growth of singular solutions. Complex-space singularities in solutions of such PDEs can be tracked by studying the temporal evolution of spectra by using the \textit{analyticity-strip technique}~\cite{sulem}. However, as the solution proceeds towards finite-time blowup, the approximations of singular solutions, obtained by using finite-resolution and finite-precision numerical methods, often suffer from loss of accuracy or stability or both. This is because the flow fields develop increasingly small-scale structures for times close to the time of blowup~\cite{gibbon2008euler}. 

The fields are approximated by employing spectral methods in space; and the governing PDEs are solved via finite-difference time-marching schemes~\cite{kerr1993evidence,cichowlas2005evolution}. The resulting numerical approximation is the exact solution of the spectrally truncated version of the governing PDE; 
but it deviates from the true solution of the PDE as the solution approaches the time of singularity (or potential singularity). These deviations first appear as localised, oscillatory structures in real space; they have been called \textit{tygers}. Tygers were first reported in Fourier pseudospectral approximations of the one-dimensional (1D) inviscid Burgers quation~\cite{ray2011resonance}, where they were attributed to nonlinear wave-particle resonances in regions of flow with positive strain. The spontaneous emergence of tygers in real space is accompanied by complex-space singularities crossing into the analyticity strip of width $\Delta_x$, where $\Delta_x$ is the smallest grid spacing in the domain. Tygers have also been reported in spectral studies of solutions of the 3D-axisymmetric, wall-bounded, incompressible Euler PDE~\cite{kolluru2022insights} and of the 3D incompressible Euler PDE~\cite{murugan2023genesis}.

Recently, the complex-time analytic structure of solutions of the 1D inviscid Burgers equation has been reinvestigated by using high-order time-Taylor-series methods~\cite{rampf}. Unlike earlier work, which concentrates on the singularity on the real time axis, Ref.~\onlinecite{rampf} charts the landscape of singularities in the complex-time plane, by investigating the convergence of time-Taylor series of the 1D Burgers velocity field at each spatial point. For single- and multi-mode sine-wave initial data, the authors of Ref.~\onlinecite{rampf} find several singularities, which are arranged in the form of an eye that is centered at the origin. Furthermore, they find that the truncated time-Taylor-series approximation of the velocity field shows spatially localised oscillatory structures, called \textit{early-time resonances}, that are reminiscent of the tygers~\cite{ray2011resonance} mentioned above. These early-time resonances emerge at the spatial points at which the convergence of the Taylor-series expansion is first lost; and the corresponding complex-time singularities, situated off of the real time axis, are the convergence-limiting singularities~\cite{rampf}.

We investigate the complex-time analytic structure of solutions of the 3D-axisymmetric, wall-bounded, incompressible Euler equations, starting from the initial data proposed in Ref.~\onlinecite{luo2014potentially}, to study a possible finite-time singularity. We use our pseudospectral Fourier-Chebyshev method~\cite{kolluru2022insights}, with quadruple-precision arithmetic to compute the time-Taylor series coefficients of the flow fields, up until a high order. The resulting approximations display early-time resonances; the initial spatial location of these structures is different from that for the tygers, which we have obtained in Ref.~\onlinecite{kolluru2022insights}. We then perform asymptotic analysis of the Taylor-series coefficients, by using the Darboux-type ratio method due to Mercer and Roberts~\cite{mercer1990centre}, to extract the location and nature of the convergence-limiting singularities. We find that these singularities are distributed around the origin, in the complex-$t^2$ plane, along two curves that resemble the shape of an eye. 
We obtain similar results for the 1D wall-approximation, of the full 3D-axisymmetric Euler equation, called the 1D HL model~\cite{luo2014potentially,choi2017finite}. We use a Fourier pseudospectral method~\cite{kolluru2022insights}, with quadruple-precision arithmetic, to compute the time-Taylor-series coefficients of the flow fields. Our work examines early-time resonances, in truncated time-Taylor expansions of solutions of these PDEs, and compares them with tygers, in Galerkin-truncated pseudospectral studies~\cite{ray2011resonance,kolluru2022insights}.

The remaining part of this paper is organized as follows. For simplicity, we first discuss the 1D HL model and obtain and analyse time-Taylor series for its solutions in Sec.~\ref{subsec:1DHL}. 
We then discuss and analyse the time-Taylor series for solutions of the 3D-axisymmetric wall-bounded incompressible Euler equations in Sec.~\ref{subsec:3DaxEuler}. 
We present our results for the 1D HL model in Sec.~\ref{subsec:Results_1D}, including the complex-singularity landscape. We present similar results for the 3D-axisymmetric wall-bounded incompressible Euler equations in Sec.~\ref{subsec:Results_3D}. Finally, in Sec.~\ref{sec:Conclusions}, we conclude with a discussion of our results. In Appendices~\ref{app:Code} and \ref{app:MR} we give, respectively,  a Mathematica code for the symbolic computation of time-Taylor coefficients for the 1D HL model and some details of the Mercer-Roberts method. 

\section{Models and Methods}
\label{sec:Models_Methods}
 A 1D wall-approximation for the full 3D-axisymmetric incompressible Euler (3DAE) equation was developed by Luo and Hou~\cite{luo2014potentially}. It is now called the 1D HL model~\cite{choi2017finite}. We discuss this 1D HL model first, in Sec.~\ref{subsubsec:Model_1D}, because it is much simpler than the full 3DAE (see below). We construct the time-Taylor series representation for 
 solutions of the 1D HL model and derive recursion relations for these time-Taylor coefficients in Sec.~\ref{subsubsec:tts_1D}. We describe the Fourier pseudospectral methods and the numerical computation of the coefficients for this model in Sec.~\ref{subsubsec:pps_1D}. In Section~\ref{subsubsec:Model_3D}, we describe the 3DAE; and we derive the recursion relations for the time-Taylor coefficients of the series for the fields in Sec.~\ref{subsubsec:tts_3D}. Fourier-Chebyshev pseudospectral methods, which we use for the computation of these coefficients, are presented in Sec.~\ref{subsubsec:pps_3D}.

\subsection{1D HL model}
\label{subsec:1DHL}
\subsubsection{Model}
\label{subsubsec:Model_1D}
The 1D HL model approximates the singular dynamics of the 3DAE [see Eq.~\eqref{eq:AxisymmetricEuler} below], at the wall at $r=1$, as discussed in Refs.~\onlinecite{luo2014potentially,choi2017finite,kolluru2022insights}. This model is 
\begin{subequations}  \label{eq:1dHLeqns}
\begin{align}
    \partial_t u + v \ \partial_z u  &= 0 , \label{eq:evol_u} \\
    \partial_t \omega + v \ \partial_z \omega &= \partial_z u, \label{eq:evol_w} \\
    \partial_z v &= \mathcal{H}[\omega]\,,\label{eq:evol_v}
\end{align}
\end{subequations}
where $\mathcal{H}[.]$ is the Hilbert transform operator; $u$, $\omega$, and $v$ are related, respectively, to the angular components of the 
velocity, vorticity, and stream function in the 3DAE (see below) at $r=1$; Eq.~\eqref{eq:evol_v} is the Biot-Savart-type law for this model. The domain is periodic in $z$ with the periodicity length $L=2\pi$. 

\subsubsection{Time-Taylor series expansion and recursion relations}
\label{subsubsec:tts_1D}
We write the time-Taylor expansions
\begin{subequations}
\begin{eqnarray}
        u(z,t) &=& \sum_{n=0}^{\infty} u_n(z) t^n\,,
    \label{eq:exp_u_1d_t}\\
    \omega(z,t)&=&\sum_{n=0}^{\infty} \omega_n(z) t^n\,,
    \label{eq:exp_omega_1d_t}\\
    v(z,t)&=&\sum_{n=0}^{\infty} v_n(z) t^n\,,
    \label{eq:exp_v_1d_t}
\end{eqnarray}
\end{subequations}

where $u_n(z),\, \omega_n(z),\,$ and $v_n(z)$ are time-Taylor coefficient of order $n$, which are functions of $z$. Hereafter, we suppress the dependence on $z$ for notational simplicity. 
To construct recursion relations for the time-Taylor coefficients, we substitute the series expansions~\eqref{eq:exp_u_1d_t}-\eqref{eq:exp_v_1d_t} in Eqs.~\eqref{eq:evol_u}-\eqref{eq:evol_v}. By comparing coefficients of terms of degree $t^{n+1}$, we obtain
\begin{subequations}  \label{eq:rec_hl1d}
\begin{align}
u_{n} &= \frac{-1}{n} \sum_{l+m = n-1} v_l \ \partial_z u_m \label{eq:rec_u},\;\; n > 0\,, \\
\omega_{n} &= \frac{-1}{n} \sum_{l+m = n-1} v_l \ \partial_z \omega_m + \frac{1}{n}  \partial_z u_{n-1},\;\; n > 0\,,\;\; {\rm{and}}\label{eq:rec_w}\\
v_{n} &= \int \mathcal{H}(\omega_n (z')) dz',\;\; n > 0\,. \label{eq:rec_v}
\end{align}
\end{subequations}
The initial conditions are
$u_0(z) = u(z,t=0)$, $\omega_0(z) = \omega(z,t=0)$, and $v_0(z) = \int \mathcal{H}(\omega_0(z')) dz' $.

Once the initial data are specified, we can use the recursion relations in Eqs.~\eqref{eq:rec_hl1d} to derive closed-form expressions for $u_n(z), \ \omega_n(z)$ and $v_n(z)$, analytically. However, the derivation of high-order coefficients quickly becomes cumbersome, even if we use symbolic-computation software [see Appendix~\ref{app:Code} for details]. Therefore, we compute the time-Taylor coefficients on a finite grid by using quadruple-precision Fourier pseudospectral numerical methods. 

\subsubsection{Fourier pseudospectral methods}
\label{subsubsec:pps_1D}
In the $L$-periodic domain, we define a uniform collocation grid of $N$ points given by $\mathbf{X}_N= \{z_j = j\Delta z \ : j=0,1,...,N-1 \}$ where $\Delta z=L/N$. The Fourier pseudospectral projection operation $P_N$ is:
\begin{subequations}
\begin{align}
    P_N u(z,t) &= \sum_{|k|\le N/2} \hat{u}(k,t) e^{i \frac{2\pi}{L} kz}\,; \\
        \hat{u}(k,t) &= \frac{1}{N}\sum_{j=0}^{N-1} u(z_j,t) e^{-i \frac{2\pi}{L}kz_j}\,. \label{eq:fourcoeff}
\end{align}
\end{subequations}

Given the initial data $u_0(z), \ \omega_0(z)$ and $v_0(z)$, we first perform a pseudospectral projection to obtain $P_N u_0$, $P_N \omega_0$, and $P_N v_0$  by using the \texttt{FFTW3} library~\cite{frigo2005design}. To compute $u_n(\mathbf{X}_N) \coloneqq \{ u_n(z) : z \in \mathbf{X}_N \}$, we set $n=1$ in Eq.~\eqref{eq:rec_u}. We evaluate the derivative $\partial_z u_n$ on the right-hand side (RHS) of Eq.~\eqref{eq:rec_u} in Fourier spectral space. The nonlinear terms are computed in real space. 
Similarly, we evaluate $\omega_1(\mathbf{X}_N)$ from Eq.~\eqref{eq:rec_w}. We use $\omega_1(\mathbf{X}_N)$ to compute $v_1(\mathbf{X}_N)$ from Eq.~\eqref{eq:rec_v}; the integration is performed in Fourier spectral space. The values of $u_1(\mathbf{X}_N)$, $\omega_1(\mathbf{X}_N)$, and $v_1(\mathbf{X}_N)$ can then be used to compute the values of $u_2(\mathbf{X}_N)$, $\omega_2(\mathbf{X}_N)$, and $v_2(\mathbf{X}_N)$. We repeat 
this procedure for $n \geq 2 $. 

\subsection{3D-axisymmetric wall-bounded incompressible Euler equation}
\label{subsec:3DaxEuler}

\subsubsection{Model}
\label{subsubsec:Model_3D}

We now consider the 3D Euler equation in the vorticity--stream function formulation:
\begin{align}
\bom_t + \bu \cdot \nabla \bom = \bom \cdot \nabla \bu \,,
\label{eq:3DEuler}
\end{align}
where $\bom = \nabla \times \bu$ is the vorticity, $\bu = \nabla \times \bpsi$ is the velocity field, and $\bpsi$ is the vector-valued stream function. The Poisson equation $\bom = - \nabla^2 \bpsi$ gives the relation between the vorticity and the stream function.\par 

To represent axisymmetric fields, we use $
\bu(r,z) = u^r (r,z) \ \mathbi{\hat{e}_r} + u^{\theta}(r,z)  \
\mathbi{\hat{e}_{\theta}} + u^z(r,z)  \ \mathbi{\hat{e}_z} $, where $ \
\mathbi{\hat{e}_r} , \ \mathbi{\hat{e}_{\theta}}$, and $ \ \mathbi{\hat{e}_z}$
are unit vectors in the cylindrical coordinate system. Furthermore, we define new variables in terms of the angular components $u^{\theta}$, $\omega^{\theta}$, and $\psi^{\theta}$:
\begin{equation}
u^1 = u^{\theta}/r; \qquad  \omega^1 = \omega^{\theta}/r; \qquad \psi^1=\psi^{\theta}/r\,.
\label{eq:u1etc}
\end{equation}

The 3D-axisymmetric wall-bounded incompressible Euler (3DAE) equations are then given in terms of the new variables~\eqref{eq:u1etc} by
\begin{subequations}
\label{eq:AxisymmetricEuler}
\begin{align}
u^1_t + u^ru^1_r + u^zu^1_z &= 2u^1\psi_z^1 ,\label{eq:main1}\\
\omega_t^1 + u^{r}\omega_r^1 + u^z\omega_z^1 &= ((u^1)^2)_z ,\label{eq:main2}\\
-\Big( \partial_r^2 + \frac{3}{r}\partial_r + \partial_z^2 \Big) \psi^{1} &= \omega^{1}, \label{eq:main3}
\end{align}
\end{subequations}
with $u^{r} = - r\psi^{1}_{z}$ and $u^{z} = 2\psi^{1}+r\psi^{1}_{r}$. For a detailed derivation, see Refs.~\onlinecite{majda,luo2014potentially}. So long as the solutions to Eq.~\eqref{eq:AxisymmetricEuler} are smooth [$C^{\infty}(\mathbf{R} \times \bar{\mathbf{R}}^{+})$], $u^{\theta}, \omega^{\theta}$, and $\psi^{\theta}$ all vanish at $r=0$; this ensures that the coordinate singularity at $r=0$ does not enter the evolution equations~\eqref{eq:AxisymmetricEuler} directly~\cite{liuwang}. Here, $\mathbf{R}$ represents the set of real numbers and $\bar{\mathbf{R}}^{+}$ the set of affinely extended positive real numbers.

We solve Eq.~\eqref{eq:AxisymmetricEuler} in the domain $
D(1,L) = \{ (r,z) : 0 \leq r \leq 1 , 0 \leq z \leq L \}$, with $L$-periodic
boundary conditions along $\hat{e}_z$, the no-flow conditions at $r=1$~\eqref{eq:noflow},
and the following pole conditions at $r=0$~\eqref{eq:polecond}:
\begin{subequations}
\begin{eqnarray}
		\psi^{1}(r=1,z,t) &=& 0\,; \label{eq:noflow} \\
	u^{1}_r(r=0,z,t)= \omega^{1}_r(r=0,z,t)
	&=&\psi^{1}_r(r=0,z,t) = 0\,.
	\label{eq:polecond}
\end{eqnarray}
\end{subequations}
The 3D-axisymmetric field variables in Eq.~\eqref{eq:AxisymmetricEuler} at $r=1$ can be approximated by the 1D-HL-model~\eqref{eq:1dHLeqns}
fields in Eq.~\eqref{eq:1dHLeqns} as discussed in Ref.~\onlinecite{luo2014potentially}:
\begin{equation}
u(z) \rightarrow (u^1)^2 (r=1,z), \qquad \omega(z) \rightarrow \omega^1(r=1,z), \qquad v(z) \rightarrow \partial_r \psi^1 (r=1,z).
\end{equation}

\subsubsection{Time-Taylor series expansion and recursion relations}
\label{subsubsec:tts_3D}

We write the time-Taylor series expansions for the fields $u^1(r,z,t),\ \omega^1(r,z,t)$, and $u^z(r,z,t)$ as  
\begin{align}
u^1(r,z,t) = \sum_{n=0}^\infty u^1_n(r,z) \ t^n\,, \qquad \omega^1(r,z,t) = \sum_{n=0}^\infty \omega^1_n(r,z) \ t^n\,,\;\;{\rm{and}} \qquad \psi^1(r,z,t) = \sum_{n=0}^\infty \psi^1_n(r,z) \ t^n\,.    
\label{eq:series_3d}
\end{align}
Similarly, we write $u^r(r,z,t) = \sum_{n=0}^\infty u^r_n(r,z,n) t^n$ and $u^z(r,z,t) = \sum_{n=0}^\infty u^z_n(r,z,n) t^n$  for the radial and axial velocities, respectively. By substituting these series expansions for the fields in Eqs.~\eqref{eq:main1}-\eqref{eq:main3} and equating the coefficients of $t^{n+1}$ on both sides, we obtain:
\begin{subequations}
\label{eq:rec_3d}
\begin{align}
        u^1_{n} (r,z) = - \frac{1}{n} \sum_{l+m=n-1} \Big( u^r_l \partial_r u^1_{m} + u^z_l \partial_z u^1_{m} - 2 u^1_l \partial_z \psi^1_{m}  \Big)\,; \label{eq:rec_3d_u} \\
        \omega^1_{n} (r,z) = - \frac{1}{n} \sum_{l+m=n-1} \Big( u^r_l \partial_r \omega^1_{m} + u^z_l \partial_z \omega^1_{m} - \partial_z ( u^1_l u^1_{m} ) \Big)\,.
        \label{eq:rec_3d_w}
\end{align}

Once we determine $\omega^1_n(r,z)$, we use the Poisson equation~\eqref{eq:main3} to obtain $\psi^1_n(r,z)$ as follows:
\begin{align}
    - \left( \partial_r^2 + \frac{3}{r} \partial_r + \partial_z^2 \right) \psi^1_n &= \omega^1_n\,; \label{eq:rec_3d_psi} \\
    \partial_r \psi^1_n (r=0,z) = 0\,; \qquad & \qquad \psi^1_n(r=1,z) = 0\,. \label{eq:rec_3d_psi_bc}
\end{align}
Here, the boundary conditions in Eqs.~\eqref{eq:noflow}-\eqref{eq:polecond} are applied to $\psi^1_n(r,z)$ in Eq.~\eqref{eq:rec_3d_psi_bc}. We can then obtain the time-Taylor coefficients for the radial and axial velocity fields:
\begin{align}
        u^r_n &= (-r \partial_z) \psi^1_n \label{eq:rec_3d_ur}\,; \\
        u^z_n &= (2+r\partial_r) \psi^1_n\,.\label{eq:rec_3d_uz}
\end{align}
\end{subequations}
The initial data are used to define $u^1_0, \, \omega^1_0$ and $\psi^1_0$ 
[cf. Subsec.~\ref{subsubsec:tts_1D} for the 1D HL model].
  For $n=1$, $u^1_1(r,z)$ and $\omega^1_1(r,z)$ can be obtained analytically, as closed-form expressions, by using Eqs.~\eqref{eq:rec_3d_u}-\eqref{eq:rec_3d_w}. However, unlike the Biot--Savart law for the 1D HL model in Eq.~\eqref{eq:evol_v}, the Poisson equation in Eqs.~\eqref{eq:rec_3d_psi}-\eqref{eq:rec_3d_psi_bc} cannot be solved analytically for $\psi^1_1(r,z)$ via symbolic computation. Thus,
we use our Fourier-Chebyshev pseudospectral scheme~\cite{kolluru2022insights} with quadruple-precision arithmetic. We describe our methods briefly in the next Subsection~\ref{subsubsec:pps_3D} [for details see Ref.~\onlinecite{kolluru2022insights}].

\subsubsection{Fourier-Chebyshev pseudospectral method for the computation of time-Taylor coefficients}
\label{subsubsec:pps_3D}

We discretize the domain by using the collocation grid $\mathbf{X}_{N,M} \coloneqq \{ (r_i,z_j) : i=1,...,N \text{ and } j=0,...,M-1 \} $, where  the axial nodes $\{z_j = \tfrac{Lj}{M}\}_{j=0,...M-1}$ are uniformly spaced. The radial nodes $\{r_i = \frac{1}{2}(1+\cos(\frac{\pi(i-0.5)}{N})\}_{i=1,...N}$, the roots of the highest-order Chebyshev polynomial $T_N(2r-1)$ in the spectral basis, are non-uniformly spaced and cluster near $r=0$ and $r=1$. The Fourier-Chebyshev pseudospectral approximation for the field $u^1(r,z,t)$ is then given by
\begin{align}
    P_{N,M} u^1(r,z,t) = \sum_{|k|<M/2} \sum_{l=0}^N \hat{u}^1(k,l,t) \ e^{ikz} \ T_l (2r-1)\,,
\end{align}
where $T_l(x)$ is the order-$l$ Chebyshev polynomial of the first kind.

We compute the derivative terms in spectral space.  
The Poisson problem in Eqs.~\eqref{eq:rec_3d_psi}-\eqref{eq:rec_3d_psi_bc} is solved by using the Fourier-Chebyshev Tau Poisson solver described in Refs.~\onlinecite{kolluru2022insights,peyret2013spectral}. The nonlinear terms are computed in real space.
Given the initial data $u^1_0(\mathbf{X}_{N,M}), \ \omega^1_0(\mathbf{X}_{N,M})$ and $\psi^1_0(\mathbf{X}_{N,M})$, we compute $u^1_1(\mathbf{X}_{N,M})$ and $\omega^1_1(\mathbf{X}_{N,M})$ from Eq.~\eqref{eq:rec_3d_u}-\eqref{eq:rec_3d_w}. 
We then solve the Poisson problem~\eqref{eq:rec_3d_psi}-\eqref{eq:rec_3d_psi_bc} for $\psi^1_1(\mathbf{X}_{N,M})$. We repeat this process to obtain the time-Taylor coefficients for $n = 2, 3,..., N_t$, where $N_t$ is the order at which we truncate the time-Taylor expansion in Eq.~\eqref{eq:series_3d}. The choice of $N_t$ plays a crucial role in ensuring that the results of our asymptotic analysis are accurate [see Sec.~\ref{subsec:Results_3D} for details]: time-Taylor coefficients, computed for $n>N_t$, suffer from errors that arise because of (a) fixed-precision arithmetic and (b) the finite number of modes retained in any implementation of the pseudospectral method. 

\section{Results}
\label{sec:Results}

We compute and examine the time-Taylor coefficients for special choices of initial data that lead to (possible) finite-time singularities~\cite{kolluru2022insights}. We then perform asymptotic analysis~\cite{vandyke,mercer1990centre,domb1957susceptibility} of these coefficients to estimate the position and the nature of the convergence-limiting singularities. 
In Section~\ref{subsec:Results_1D}, we chart out the pattern of these singularities for a singular solution of the 1D HL model~\eqref{eq:1dHLeqns}. In Section~\ref{subsec:Results_3D}, we repeat the analysis for the potentially singular solution of the 3DAE equation~\eqref{eq:AxisymmetricEuler} with the initial condition used in Refs.~\onlinecite{luo2014potentially,kolluru2022insights,barkley,hertel}.

\subsection{1D HL-model}
\label{subsec:Results_1D}
We start with the following initial data:
\begin{align}
    u_0 (z) = \sin^2(z)\,; \qquad \omega_0(z) = 0.  
    \label{eq:IC_1DHL}
\end{align}
It has been proved~\cite{choi2017finite} that the flow, which evolves from Eq.~\eqref{eq:IC_1DHL}, develops a finite-time singularity. For a scaled version of the initial data used in Ref.~\onlinecite{kolluru2022insights}, the time of singularity lies near $t_* \simeq 0.0035$. 

\begin{table}[htbp] 
\centering
\begin{center}
$
\begin{array}{|c|c|c|c|}
\hline
{n} & u_n(z) & {\omega_n}(z) & {v}_n(z) \\
\hline 
& & & \\
 0 & \sin ^2(z) & 0 & 0 \\
 1 & 0 & \sin (2 z) & \sin (z) (-\cos (z)) \\
 2 & \sin ^2(z) \cos ^2(z) & 0 & 0 \\
 3 & 0 & \frac{1}{3} \sin (4 z) & -\frac{1}{12} \sin (4 z) \\
 4 & \frac{1}{12} \sin (2 z) \sin (4 z) & 0 & 0 \\
 5 & 0 & \frac{1}{15} (2 \sin (6 z)-\sin (2 z)) & \frac{1}{90} (3 \sin (2 z)-2 \sin (6 z)) \\
 6 & \frac{1}{270} \sin ^2(2 z) (17 \cos (4 z)+7) & 0 & 0 \\
 7 & 0 & \frac{109 \sin (8 z)-86 \sin (4 z)}{1890} & \frac{172 \sin (4 z)-109 \sin (8 z)}{15120} \\
 . & . & . & . \\ 
 . & . & . & .  \\ 
\hline 
\end{array}
$
\end{center}
\caption{Closed-form symbolic expressions of the time-Taylor coefficients $u_n(z),\omega_n(z)$, and $v_n(z)$ for low-order $n\le 7$, obtained by using the recursion relations in Eq.~\eqref{eq:rec_hl1d} 
and starting from the initial data in Eq.~\eqref{eq:IC_1DHL} [see Appendix~\ref{app:Code} for details].}
\label{tab:coeffs_hl1d}
\end{table}

\begin{figure}[htbp]
    \centering
 \centering
        \begin{tikzpicture}
 	\draw (0,0) node[inner sep=0]{\includegraphics[height=0.245\linewidth,trim={0cm 0 2.7cm 0},clip]{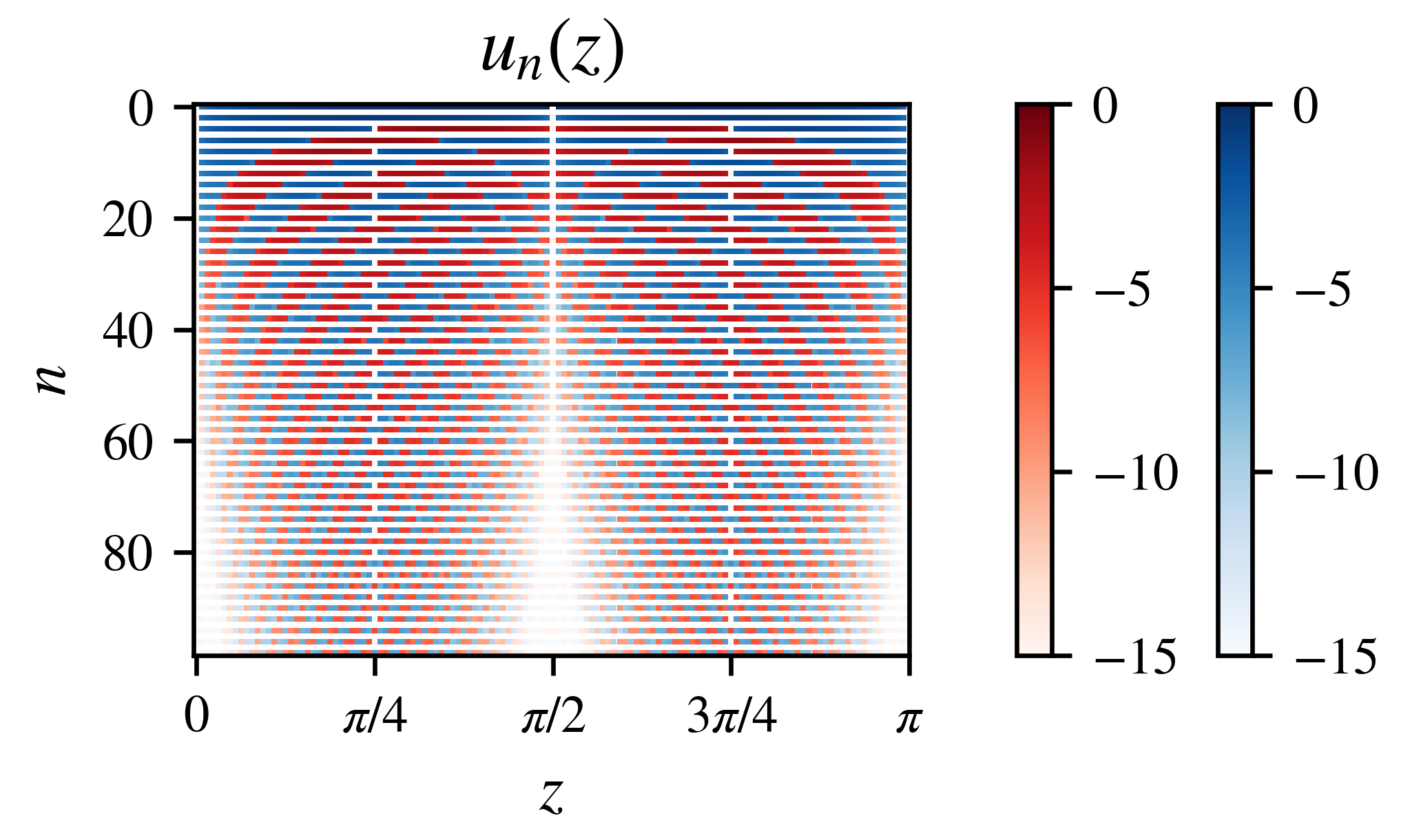}};
		\draw (-0.4,1.74) node {(a)};
	\end{tikzpicture}
         \begin{tikzpicture}
 	\draw (0,0) node[inner sep=0]{\includegraphics[height=0.245\linewidth,trim={0cm 0 2.7cm 0},clip]{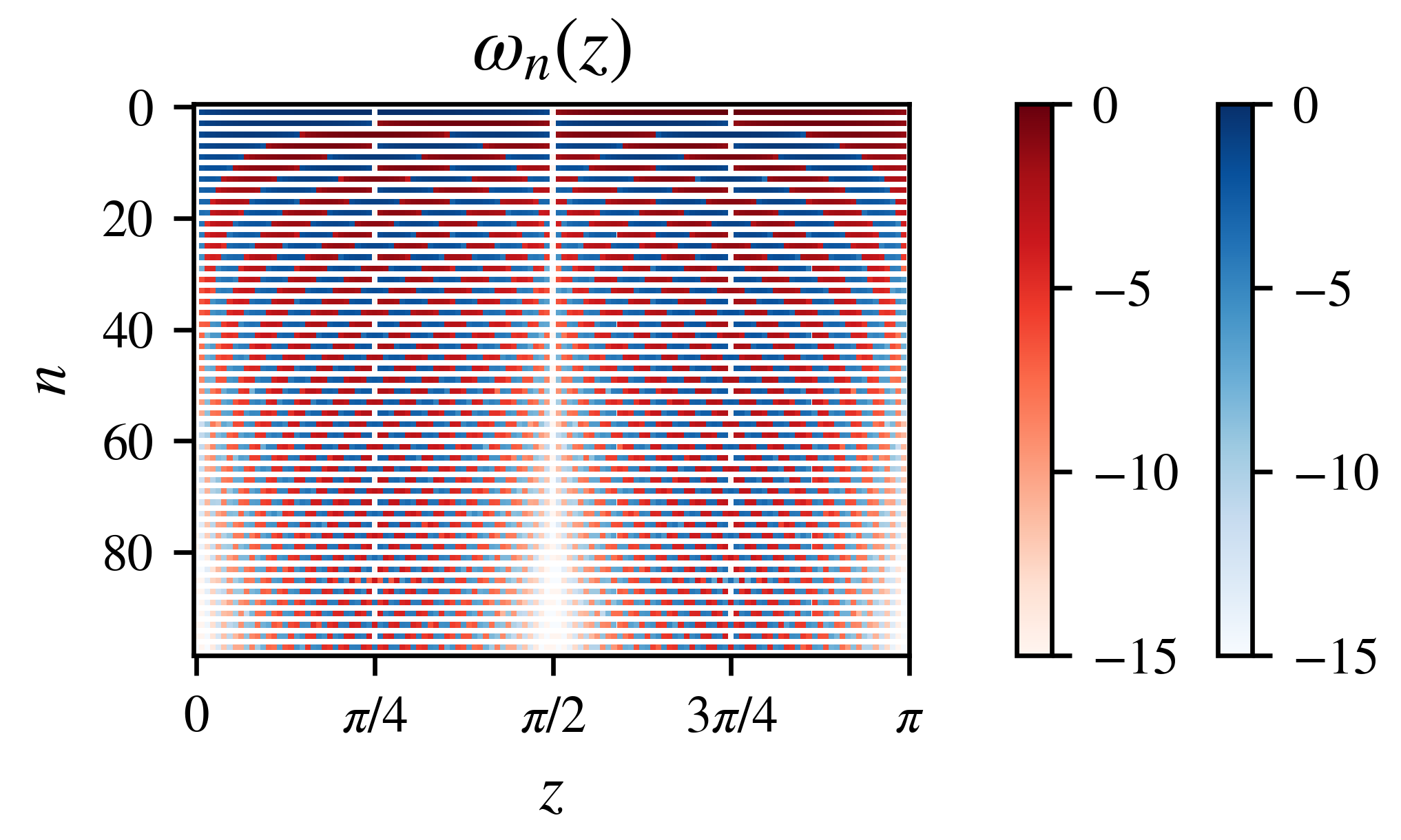}};
		\draw (-0.4,1.74) node {(b)};
	\end{tikzpicture}
         \begin{tikzpicture}
 	\draw (0,0) node[inner sep=0]{\includegraphics[height=0.245\linewidth]{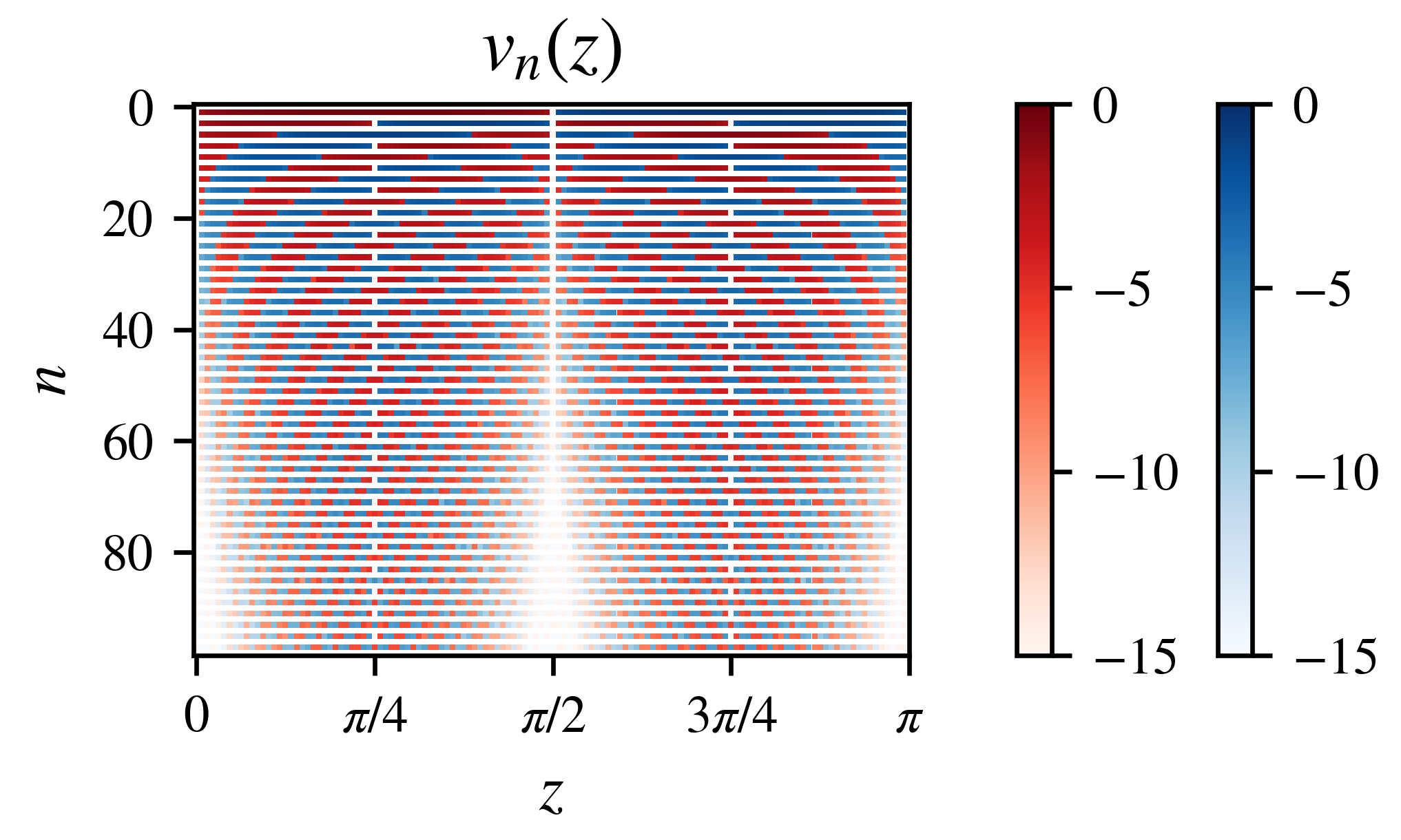}};
		\draw (-1.35,1.74) node {(c)};
	\end{tikzpicture}
    \caption{ Sign-coded heat maps of the absolute values of the time-Taylor coefficients (a) $u_n(z)$, (b) $\omega_n(z)$, and (c) $v_n(z)$, for $n\le100$ evaluated at $z \in \mathbf{X}_{N=256}$. The coefficients with positive (negative) signs are shown in red (blue); $u_n(z)$ vanishes identically for odd $n$ (white bands); $\omega_n(z)$ and $v_n(z)$ vanish identically for even $n$, because of the symmetry in the initial condition. We use a log scale (base 10) for the color bar.}
    \label{fig:1d_coeffs}
\end{figure}
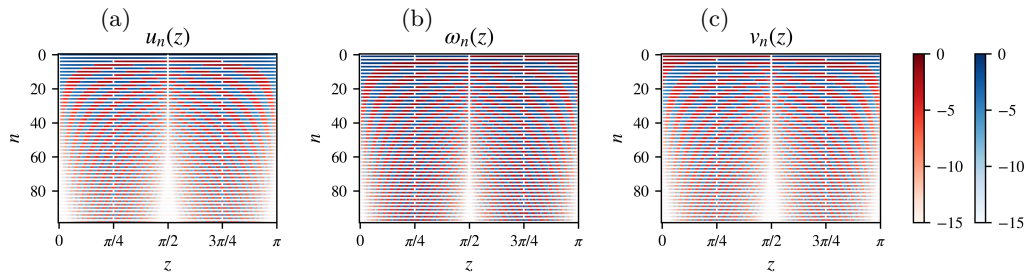

We now apply the recursion relations in Eqs.~\eqref{eq:rec_u}-~\eqref{eq:rec_v}, starting from the initial data in Eq.~\eqref{eq:IC_1DHL}. We use symbolic computation to compute the closed-form expressions, for low-order ($n \leq 7$) coefficients
$u_n(z),\ \omega_n(z)$ and $v_n(z)$, that we list 
in TABLE~\ref{tab:coeffs_hl1d}. Appendix~\ref{app:Code}
contains our Mathematica code for the determination of 
these coefficients.
The odd-order coefficients for $u(z)$ and the even-order coefficients for $\omega(z)$ and $v(z)$ vanish identically because of the symmetry of the initial data~\eqref{eq:IC_1DHL}. The difficulty of symbolic computation and the storage of high-order coefficients increases rapidly with the order $n$. Therefore, we use the Fourier pseudospectral methods outlined in Sec.~\ref{subsubsec:pps_1D} to compute the time-Taylor coefficients up to order $n \leq N_t=100$, on a uniform grid with $N=256$ points. For accuracy, it is
important to use quadruple-precision arithmetic.
{In FIG.~\ref{fig:1d_coeffs}, we present a sign-coded heat map of the absolute values of the coefficients $u_n(z), \ \omega_n(z),$ and $\psi_n(z)$ for $n \leq 100$ evaluated at the collocation points $z \in \mathbf{X}_{N=256}$; the coefficients with positive (negative) signs are shown in red (blue); the color bars use a log scale (base 10). White regions appear in these plots where the coefficients vanish identically [e.g., for odd orders in $u_n(z)$] or fall below $10^{-15}$.}

\begin{figure}[htbp]
    \centering
    \begin{tikzpicture}
    \draw (0,0) node[inner sep=0]{\includegraphics[width=\linewidth]{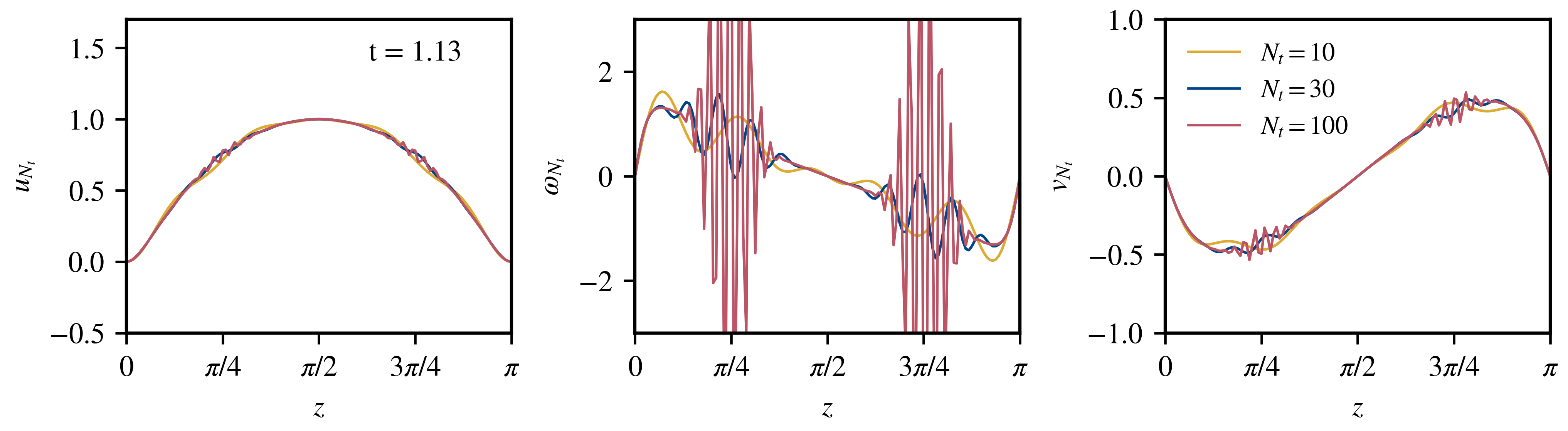}};
    \draw (-3.1,-0.8) node {(a)};
    \draw (2.1,-0.8) node {(b)};
    \draw (7.5,-0.8) node {(c)};
    \end{tikzpicture}
    \caption{Plots vs. $z$ of the truncated time-Taylor series approximations for the fields (a) $u_{N_t}(z,t)$, (b) $\omega_{N_t}(z,t)$, and (c) $v_{N_t}(z,t)$ at time $t=1.13$, summed for $N_t = 10 \text{ (yellow)}, \ 30$ (blue) and $100$ (red). Early-time resonances, which occur in all three fields, are oscillatory structures localised around $z = \{ \tfrac{\pi}{4}, \tfrac{3\pi}{4} \}$; their amplitude is largest for $\omega_{N_t}$.} 
    \label{fig:1d_slice}
\end{figure}
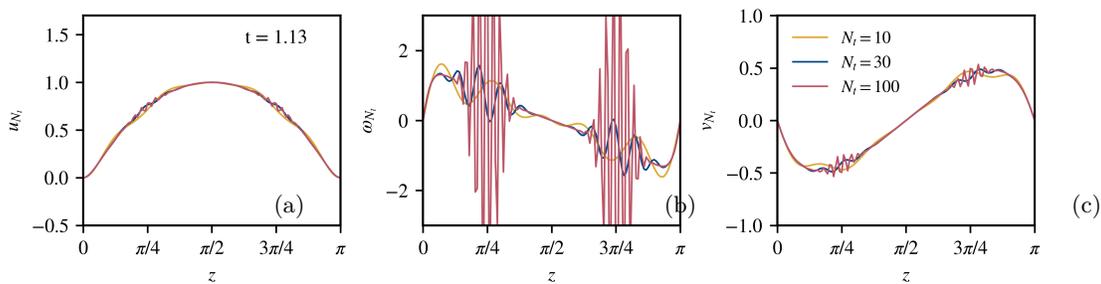

In FIG.~\ref{fig:1d_slice}, we plot the truncated series approximations $u_{N_t} \coloneqq \sum_{n=0}^{N_t} {u}_n t^n$, $\omega_{N_t} \coloneqq \sum_{n=0}^{N_t} {\omega}_n t^n$ and $v_{N_t} \coloneqq \sum_{n=0}^{N_t} {v}_n t^n$ at time $t=1.13$, for $N_t = 10 \text{ (yellow)} ,30$ (blue), and $100$ (red). 
We observe the emergence of localised oscillatory structures at $z= \{ \tfrac{\pi}{4},\tfrac{3\pi}{4},\tfrac{5\pi}{4}, \tfrac{7\pi}{4} \}$, for all the fields; we refer to these structures as \textit{early-time resonances}, a term introduced in Ref.~\onlinecite{rampf} for the 1D inviscid Burgers equation in a periodic domain.
 Similar oscillatory structures were also reported for the Fourier-pseudospectral study of the 1D HL model in Ref.~\onlinecite{kolluru2022insights}, near the time of singularity; they were identified as nonlinear wave-particle resonances called \textit{tygers}. 
However, the spatial locations of the tygers, observed in Ref.~\onlinecite{kolluru2022insights} for $u(z,t)$ and $\omega(z,t)$, are different from what we see in FIG.~\ref{fig:1d_slice}: In particular, for the initial condition~\eqref{eq:IC_1DHL} the Fourier-pseudospectral study in Ref.~\onlinecite{kolluru2022insights} yields tygers at $z= \{ \tfrac{\pi}{2}, \tfrac{3\pi}{2} \}$ for $u(z,t)$ and $\omega(z,t)$. This dissimilarity between early-time resonances and tygers, which we observe here, generalises significantly its first observation in Ref.~\onlinecite{rampf} for the 1D inviscid Burgers equation. 

In FIG.~\ref{fig:1d_slice}, the emergence of early-time resonances signals the loss of convergence of the Taylor-series approximation. For an analytic function $u(t)$, the convergence of the Taylor series about a point $t=t_0$ is limited by the nearest singularity in the complex-$t$ plane. If $R$ is the distance of the nearest singularity from the point of expansion $t_0$, 
the Taylor series loses convergence for $|t-t_0|>R$. The nature and position of this singularity (or singularities) governs the large-$n$ behaviour of the Taylor coefficients. Conversely, we can use the large-$n$ behaviour of $u_n(z), \ \omega_n(z)$, and $v_n(z)$ in FIG.~\ref{fig:1d_coeffs} to chart the positions of complex-time singularities. 

Given our initial data~\eqref{eq:IC_1DHL}, the time-Taylor series expansion for $u(z,t)$  [TABLE~\ref{tab:coeffs_hl1d}] contains only even-order terms, so we use $q \equiv t^2$ as the expansion variable. In contrast, the expansions for $\omega(z,t)$ and $v(z,t)$ do not have even-order terms in $t$; this allows us to rewrite their series in terms of $q \equiv t^2$, with a prefactor of $t=q^{1/2}$ as follows:
\begin{subequations}
\label{eq:lh1d_t2exp}
\begin{eqnarray}
    u_{N_t}(z,q) &=& \sum_{n=0}^{N_t/2} u_{2n}(z) {q}^n\,;\;\; \label{eq:lh1d_t2exp_u} \\
    \omega_{N_t}(z,q) &=& q^{1/2} \ \sum_{n=0}^{N_t/2} \omega_{2n+1}(z) {q}^n\,;\;\;\label{eq:lh1d_t2exp_w}  \\
    v_{N_t}(z,q) &=& q^{1/2} \ \sum_{n=0}^{N_t/2} v_{2n+1}(z) {q}^n\,. \label{eq:lh1d_t2exp_v} 
\end{eqnarray}
\end{subequations}

Furthermore, in FIG.~\ref{fig:1d_coeffs} we observe that $u_{2n}(z=0)$ and $u_{2n}(z=\pi)$ display alternating signs with a long period; 
the period over which the sign of the coefficient alternates is reduced as $z\rightarrow\{ \pi/2 \}$. Given these alternating signs, we cannot use the standard Domb--Sykes ratio method~\cite{domb1957susceptibility,vandyke} for determining the radius of convergence of the series $u_{N_t}(z,t)$. This method can only be applied when the coefficients are all of the same sign or have strictly alternating signs. When the signs alternate with a period greater than $1$, the sign of the ratio $u_{2(n+1)}(z)/u_{2n}(z)$ changes with $n$, so $u_{2(n+1)}(z)/u_{2n}(z)$ does not have a well-defined limit as $n\to\infty$. 

\begin{figure}[htbp]
    \centering
    \begin{tikzpicture}
    \draw (0,0) node[inner sep=0]{\includegraphics[width=0.325\linewidth]{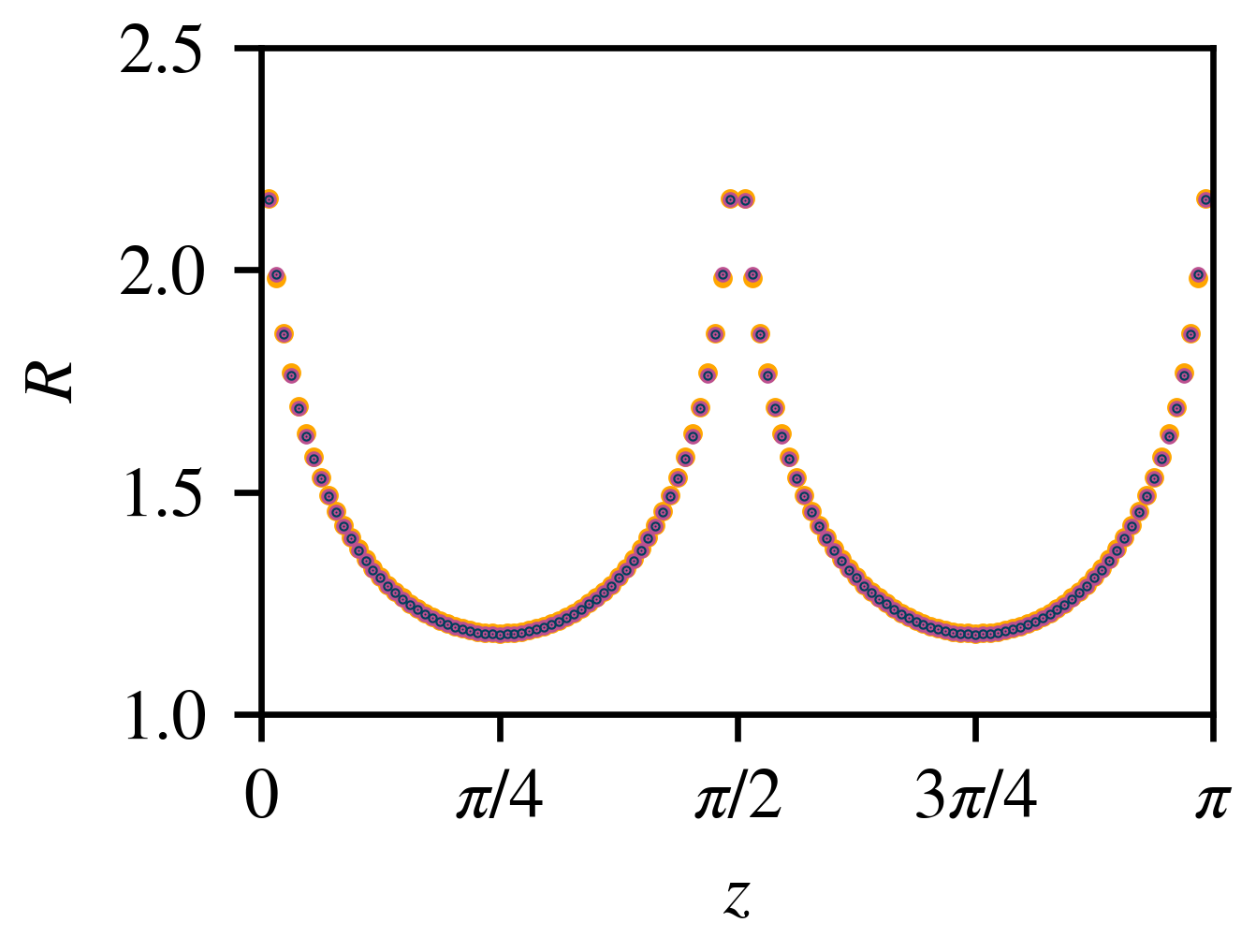}};
    \draw (-1.1,1.45) node {(a)};
    \end{tikzpicture}
    \begin{tikzpicture}
    \draw (0,0) node[inner sep=0]{\includegraphics[width=0.325\linewidth]{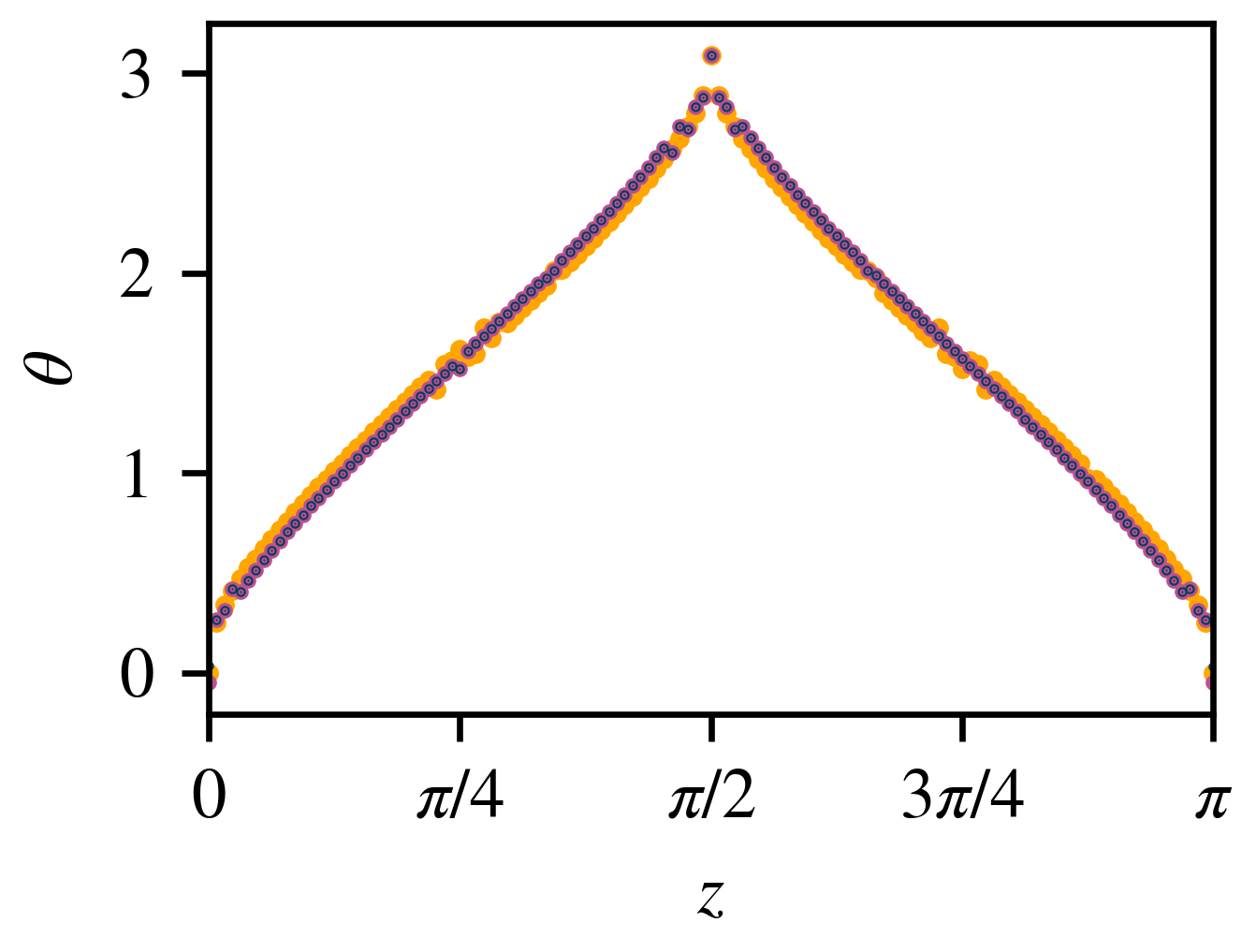}};
    \draw (-1.2,1.45) node {(b)};
    \end{tikzpicture}
    \begin{tikzpicture}
    \draw (0,0) node[inner sep=0]{\includegraphics[width=0.325\linewidth]{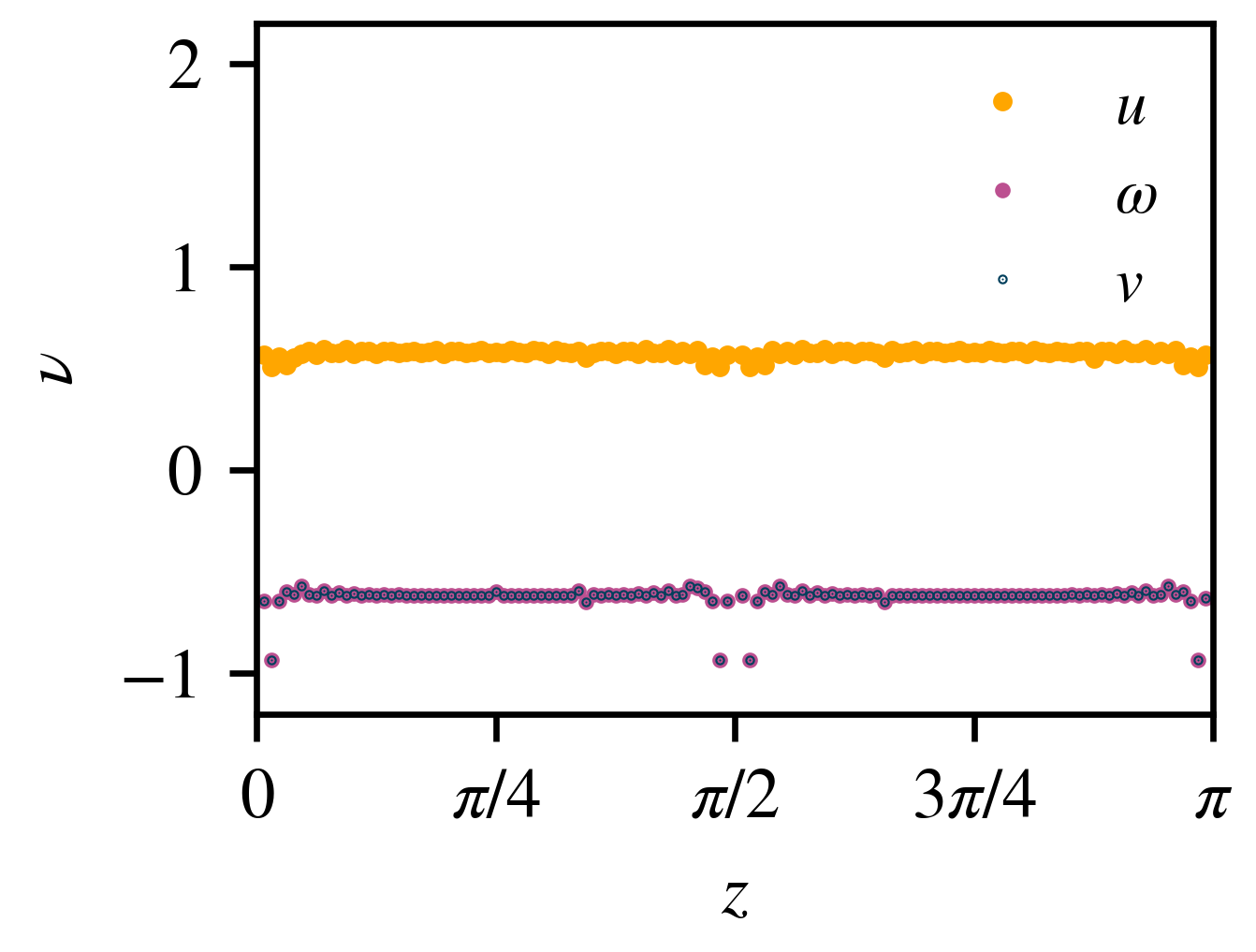}};
    \draw (-1.1,1.45) node {(c)};
    \end{tikzpicture}
    \caption{Plots vs. $z$ of the estimates of (a) $R$, (b) $\theta$, and (c) $\nu$, obtained from $u_n \text{(orange)}, \ \omega_n$(pink) and $v_n$(blue), for $z \in \mathbf{X}_{256}$ and $n\le N_t=100$, by using the MR method. A full nonlinear fit of $B_k^2$ in Eq.~\eqref{eq:MR_coeff_text} is performed by using the LMFIT package in Python.}
    \label{fig:1d_rthetanu} 
\end{figure}
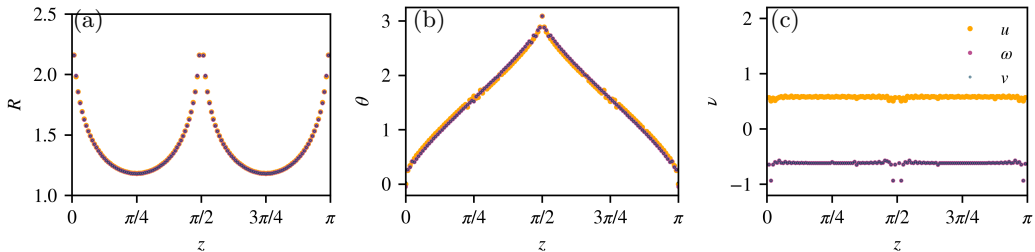

Mercer and Roberts~\cite{mercer1990centre} (MR) generalised the Domb--Sykes method for series with coefficients whose signs change with order $n$ and have a period of alternation greater than $1$. In the MR method, the pattern of alternating signs is attributed to a pair of complex-conjugate singularities that lie close to the real domain [see Appendix~\ref{app:MR}]. For a given value of $z$, the Taylor series $u_{N_t} (z,q)$ is approximated by the MR-model function $\mathfrak{u}(z,\mathbf{q})$, in complex-$\mathbf{q}$ space, where $\mathbf{q}$ is the complexified version of the variable $q=t^2$:
\begin{align}
    \mathfrak{u}(z,\mathbf{q}) = \left( 1-\frac{\mathbf{q}}{\mathbf{q_*}}\right)^{\nu(z)} + \left(1-\frac{\mathbf{q}}{\overline{\mathbf{q}}_*}\right)^{\nu(z)}\,; \qquad \mathbf{q_*}(z) \coloneqq R(z)e^{i\theta(z)}.
    \label{eq:MR_1d}    
\end{align}
Here, $\mathbf{q}_*$ denotes the position of the convergence-limiting complex singularity and $\overline{\mathbf{q}}_*$ its complex conjugate. The unknown functions $ R(z) > 0, \ \theta(z) \in [0,\pi]$, with $\nu(z) \in \mathbb{R}$ neither zero nor a positive integer. 
For each value of $z \in \mathbf{X}_N $, we construct the MR coefficients $B_k^2(z)$ given in Eq.~\eqref{eq:MR_coeff_text}. We then perform a nonlinear fit for $B^2_k$ with the functional forms given in Eq.~\eqref{eq:modfn_ts_text} below: 
\begin{subequations}
	\begin{align}
		B^2_k (z) &= \frac{u_{k+1}(z) \cdot u_{k-1}(z) - u^2_k(z)}{u_k(z) \cdot u_{k-2}(z) - u^2_{k-1}(z) }\,; \qquad 2 \leq k \leq (N_t - 1) 	\label{eq:MR_coeff_text} \,; \\
	u_n(z) &= 2(-1)^n \binom{\nu(z)}{n} R(z)^{-n} \cos(n \theta(z)) \,; \qquad 0\leq n \leq N_t\,. \label{eq:modfn_ts_text}
\end{align}
\end{subequations}
From these we obtain estimates for the functions $R(z), \ \theta(z),$ and $\nu(z)$ [FIGs.~\ref{fig:1d_rthetanu} (a)-(c), respectively], for $z \in \mathbf{X}_{256}$, by using the LMFIT package~\cite{newville2016lmfit} in Python. In Appendix~\ref{app:MR}, we discuss the 
$\tfrac{1}{k}$ linear fit for $B_k (z)$; this linear fit works for $z=\{ 0, \, \tfrac{\pi}{8} \, \tfrac{\pi}{4} \}$; for other values of $z$ we must, perforce, use the nonlinear fit provided by LMFIT. The estimates from Eqs.~\eqref{eq:lh1d_t2exp_u}, ~\eqref{eq:lh1d_t2exp_w}, and ~\eqref{eq:lh1d_t2exp_v} are shown, respectively, in yellow, pink, and blue in FIGs.~\ref{fig:1d_rthetanu} (a)-(c). 

\begin{figure}[htbp]
    \centering    
    \includegraphics[width=0.354\linewidth]{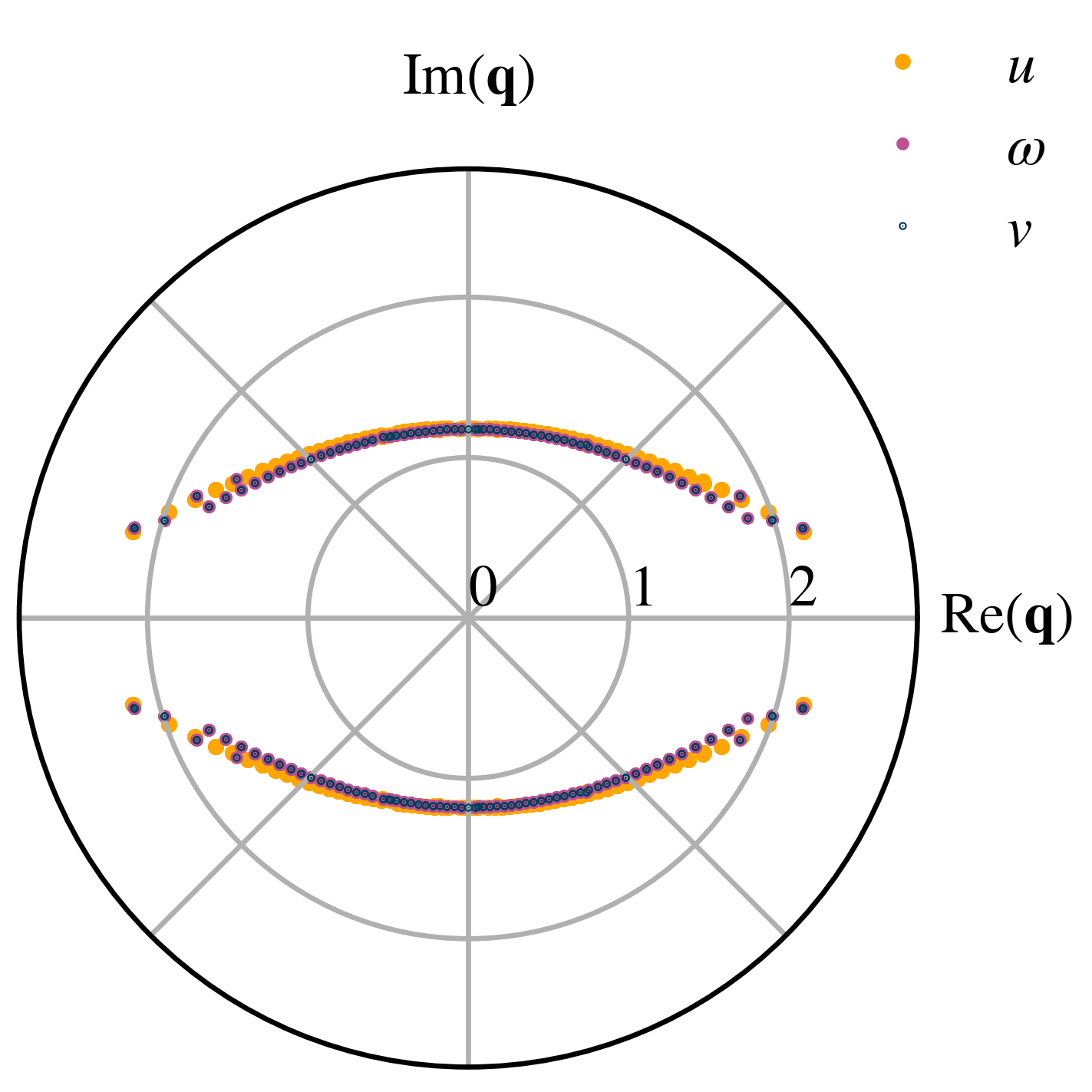}
    \caption{Polar plot, in the complex-$\mathbf{q}$ plane, of $R(z)$ vs. $\theta(z)$, depicting the positions of convergence-limiting complex singularities $\mathbf{q}_* = Re^{i \theta}$, as well as their complex-conjugates, for $u$ (yellow), $\omega$ (pink), and $v$ (blue) at $z \in \mathbf{X}_{N=256}$. The singularities are arranged in the shape of an eye, which is squashed along the $\Im(\mathbf{q})$ axis and is centered at the origin.} 
    \label{fig:1d_eye_q}
\end{figure}

The convergence-limiting singularities for $z=\{0,\pi\}$ occur at $\theta(0)=\theta(\pi)=0$; and for $z=\tfrac{\pi}{2}$ they occur at $\theta(\tfrac{\pi}{2}) = \pi$ [panel (b) of FIG.~\ref{fig:1d_rthetanu}]; i.e., these singularities are on the positive and negative real axes in the complex-$\mathbf{q}$ plane, respectively. They lie farthest from the origin because the radius of convergence $R(z)$ assumes its maximum value at these points [panel (a) of FIG.~\ref{fig:1d_rthetanu}]; $R(z)$ assumes its minimum value $R_m$ at $z = \{ \tfrac{\pi}{4} ,\tfrac{3 \pi}{4} \}$; here, $\theta (z) = \tfrac{\pi}{2}$. Thus, singularities for $z = \{ \tfrac{\pi}{4}, \tfrac{3 \pi}{4} \}$ are closest to the origin and are positioned on the imaginary axis $\Re(\mathbf{q})=0$. In particular, when $t^2> R_{m}$, where $R_m = R(z=\{ \tfrac{\pi}{4}, \tfrac{3\pi}{4} \})$, the fields in FIG.~\ref{fig:1d_slice} develop early-time resonances. The minimum radius of convergence $R_m \simeq 1.17$ or $t_m \simeq 1.08$ for all fields [see FIG.~\ref{fig:1d_rthetanu} and FIG.~\ref{fig:1d_eye_q}];
{however, early-time resonances appear more prominently in $\omega_{N_t}$ than in $u_{N_t}$ and $v_{N_t}$ [see FIG.~\ref{fig:1d_slice}].}

In FIG.~\ref{fig:1d_eye_q}, we plot $(R(z),\theta(z))$, for different values of $z$ in the complex $\mathbf{q}$ plane, by using polar coordinates. These singularities form of an approximately elliptical eye, which is centered at the origin and compressed along the $\Im(\mathbf{q})$ axis. 
The arrangement of such complex singularities in the form of an eye was first reported for time-Taylor expansions of the 1D inviscid Burgers equation~\cite{rampf}, where the eye occurs  in the complex-$t$ plane. Given the symmetries of initial conditions that we use, these eyes occur in the complex-$\mathbf{q}$ plane [see FIG.~\ref{fig:1d_eye_q}]. 
{In panel (c) of FIG.~\ref{fig:1d_rthetanu}, we plot the 
{exponent} $\nu(z)$ of the convergence-limiting singularities for all the fields; here, we observe that $\nu$ takes {non-integer} values that depend on $z$. This $z$ dependence is significant only near $z=\{0,\tfrac{\pi}{2},\pi\}$; otherwise $\nu \simeq \tfrac{1}{2}$ for $u$, and $\nu=-\tfrac{1}{2}$ for $\omega$ and $v$. }
{We cannot obtain accurate estimates for singularities that lie arbitrarily close to the real axis in FIG.~\ref{fig:1d_eye_q}. This is because the MR method is effective only when the series has settled down to its final regular behaviour. Given finite-resolution and finite-precision errors, the truncated Taylor series that we use in Eq.~\eqref{eq:lh1d_t2exp} does not contain enough terms to capture this behaviour. [See Appendix~\ref{app:MR} for details.] }

\subsection{3D-axisymmetric wall-bounded incompressible Euler equation}
\label{subsec:Results_3D}

We use the singular initial condition proposed by Ref.~\onlinecite{luo2014potentially} for this model to obtain
\begin{subequations}
  \label{eq:3d_ic}
\begin{align}
    u^1_0(r,z) = e^{-30(1-r^2)^4} \sin{\left(\tfrac{2\pi}{L}z \right)}\,, \\
    \omega^1_0(r,z)= 0\,, \qquad {\rm{and}}\qquad \psi^1_0(r,z) = 0\,,  
\end{align}
\end{subequations}
where $L=2\pi$ \footnote{Note that Refs.~\onlinecite{luo2014potentially,kolluru2022insights} use $L=1/6$.}.
For every point $(r,z) \in \mathbf{X}_{N,M}$, we use our Fourier-Chebyshev pseudospectral method~\cite{kolluru2022insights} to compute  $u^1_n(\mathbf{X}_{N,M})$ and $\omega^1_n (\mathbf{X}_{N,M})$ via the recursion relations~\eqref{eq:rec_3d_u} and \eqref{eq:rec_3d_w}, respectively. We then solve the Poisson problem~\eqref{eq:rec_3d_psi}-\eqref{eq:rec_3d_psi_bc} for $\psi^1_n(\mathbf{X}_{N,M})$ by using the Tau Poisson solver in Fourier-Chebyshev spectral space~\cite{kolluru2022insights}. The time-Taylor coefficients for the radial and axial velocities, $u^r_n(\mathbf{X}_{N,M})$ and $u^z_n(\mathbf{X}_{N,M})$, are then computed from $\psi^1_n(\mathbf{X}_{N,M})$ [see Sec.~\ref{subsubsec:tts_3D}]. 

\begin{figure}[htbp]
    \centering
 \centering
        \begin{tikzpicture}
 	\draw (0,0) node[inner sep=0]{\includegraphics[height=0.244\linewidth,trim={0cm 0 2.5cm 0},clip]{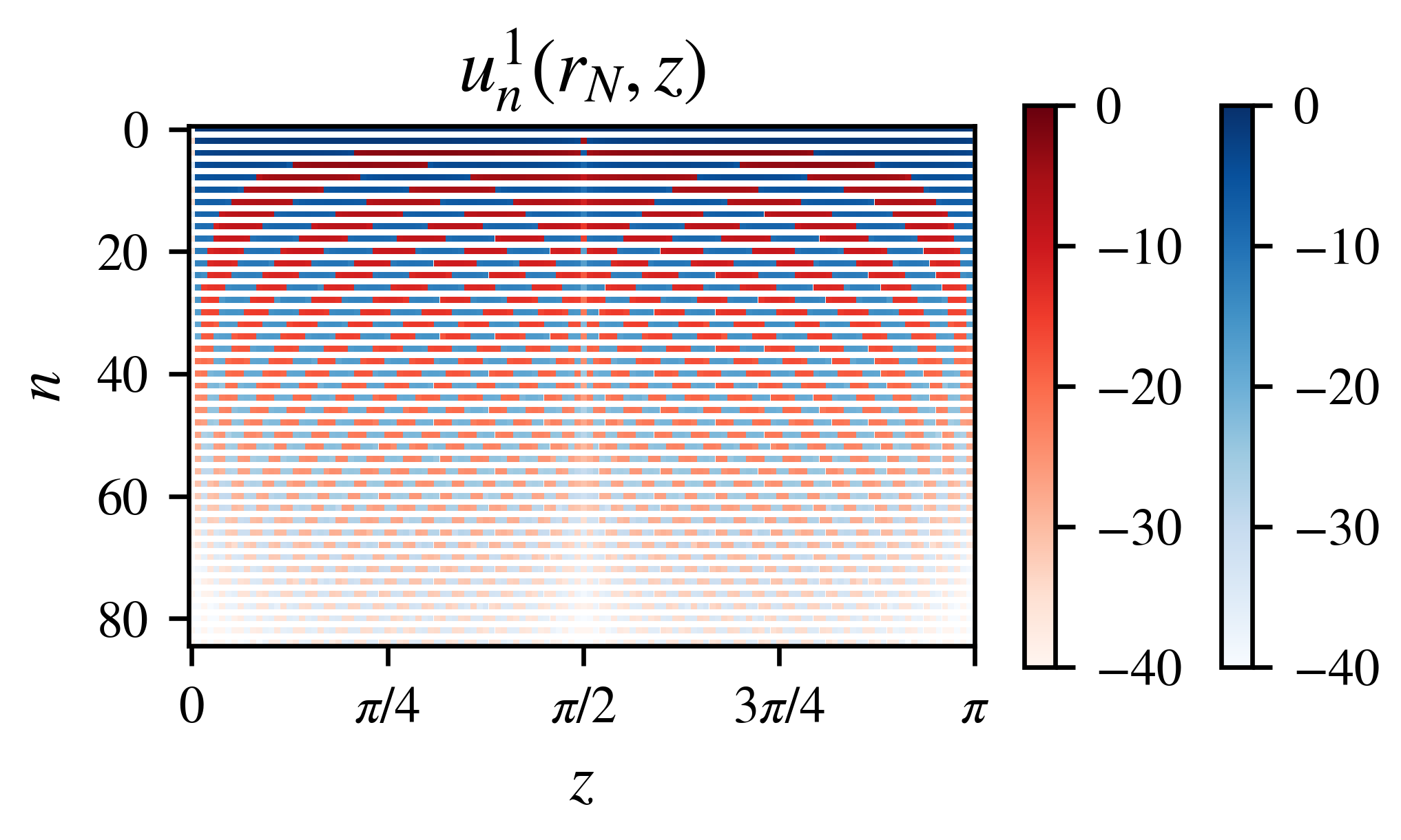}};
		\draw (-0.8,1.66) node {(a)};
	\end{tikzpicture}
         \begin{tikzpicture}
 	\draw (0,0) node[inner sep=0]{\includegraphics[height=0.244\linewidth,trim={0cm 0 2.5cm 0},clip]{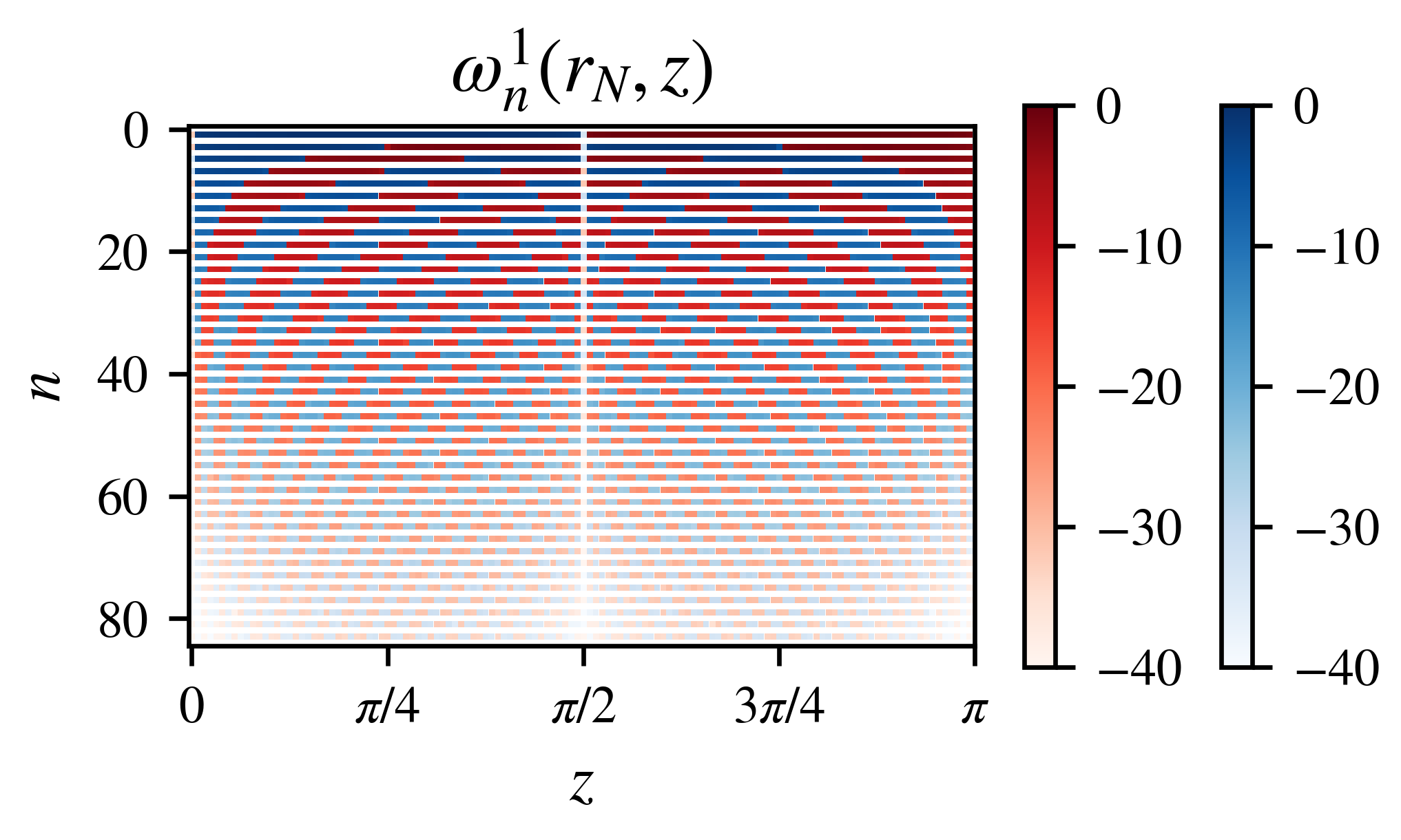}};
		\draw (-0.8,1.66) node {(b)};
	\end{tikzpicture}
         \begin{tikzpicture}
 	\draw (0,0) node[inner sep=0]{\includegraphics[height=0.244\linewidth]{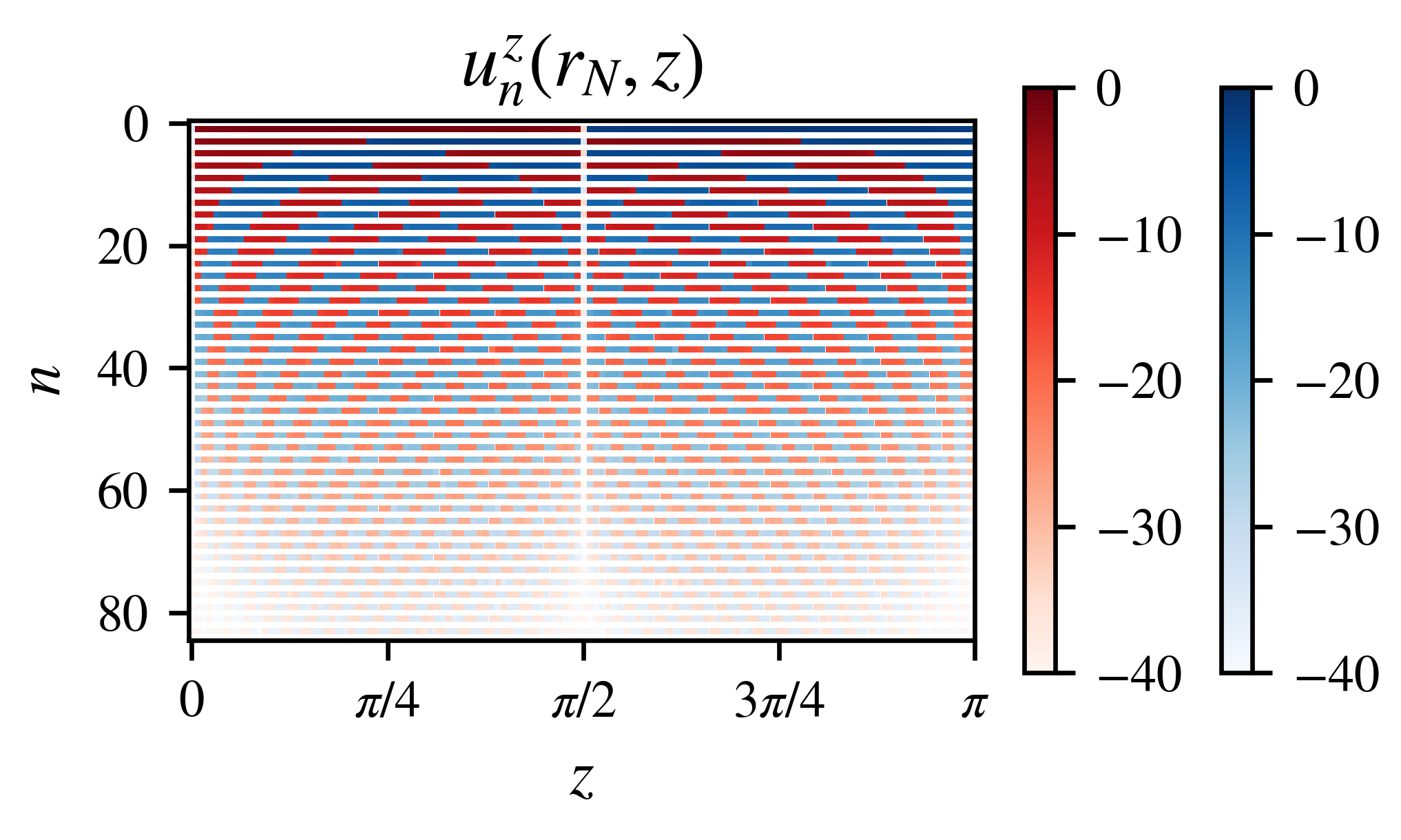}};
		\draw (-1.65,1.66) node {(c)};
	\end{tikzpicture}
    \caption{
     Sign-coded heat maps of the absolute values of the time-Taylor coefficients (a) $u^1_n(r_N,z)$, (b) $\omega^1_n(r_N,z)$, and (c) $u^z_n(r_N,z)$, for $n\le N_t=86$ and evaluated  at $ r_N, \ z \in \mathbf{X}_{N,M}$, where $(N=128,M=256)$. The coefficients with positive (negative) signs are shown in red (blue);  $u^1_n(r_N,z)$ vanish identically for odd $n$ (white bands); and $\omega^1_n(r_N,z)$ and $u^z(r_N,z)$ vanish identically for even $n$, because of the symmetry of the initial data. We use a log scale (base 10) for the color bar.}
    \label{fig:3d_coeffs}
\end{figure}
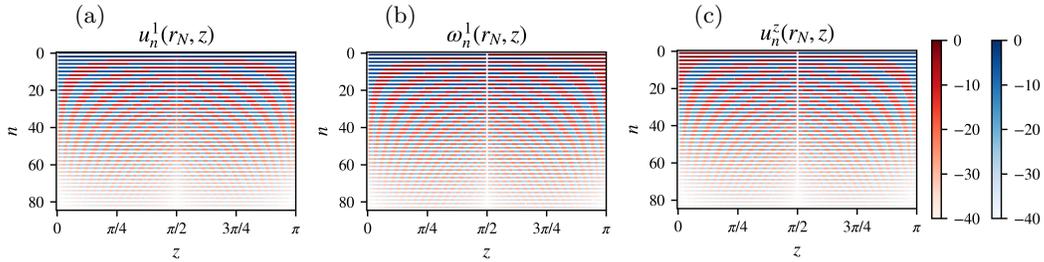

In FIG.~\ref{fig:3d_coeffs}, we present the sign-coded heat map of the absolute values of the coefficients $u^1_n, \, \omega^1_n$ and $\psi^1_n$ for $n \leq 86$ evaluated at the collocation points $(r_N, z) \in \mathbf{X}_{N,M}$ where $(N=128,M=256)$; $r_N$ is the Chebyshev node that lies closest to the boundary at $r=1$.
As in FIG.~\ref{fig:1d_coeffs}, the coefficients with positive (negative) signs are shown in red (blue), with log scale (base 10) color bars; the coefficients vanish identically [e.g., for odd orders in $u^1_n$] or fall below $10^{-40}$ in white regions. The odd-order coefficients for $u^1_n$ and the even-order coefficients for $\omega^1_n$ and $u^z_n$ vanish identically because of the symmetry of the initial condition~\eqref{eq:3d_ic}. 

\begin{figure}[htbp]
   \centering
    \begin{tikzpicture}
    \draw (0,0) node[inner sep=0]{\includegraphics[width=\linewidth]{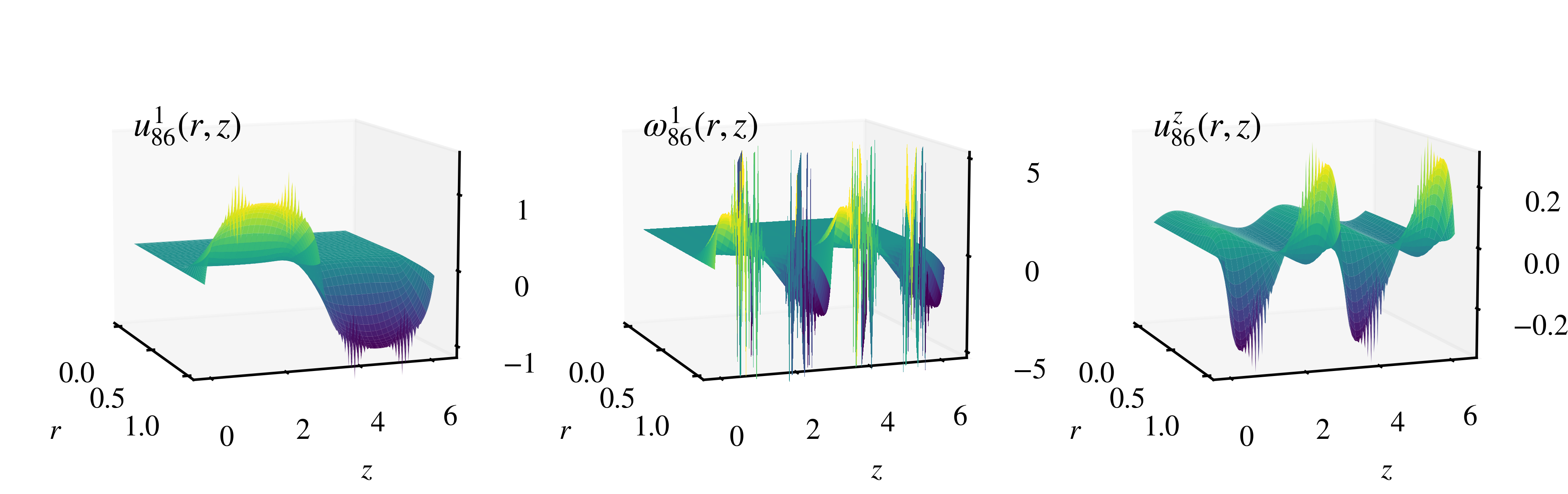}};
    \draw (-3.7,1.4) node {(a)};
    \draw (1.4,1.4) node {(b)};
    \draw (6.8,1.4) node {(c)};
    \end{tikzpicture}
   \caption{ Surface plots vs. $(r,z)$ of the time-Taylor series approximations of the fields (a) $u^1$, (b) $\omega^1$, and (c) $u^z$, truncated at $N_t=86$ and evaluated at time $t=2.6$ on a Fourier-Chebyshev collocation grid $\mathbf{X}_{N,M}$, where $(N=128,M=256)$. We see the development of early-time resonances in all fields near the wall at $r=1$. These resonant oscillations are localised around $z = \{ \tfrac{\pi}{4},\tfrac{3\pi}{4}, \tfrac{5\pi}{4},\tfrac{7\pi}{4} \}$; they grow in amplitude and spread outwards in time.}
   \label{fig:3d_Sum}
\end{figure}
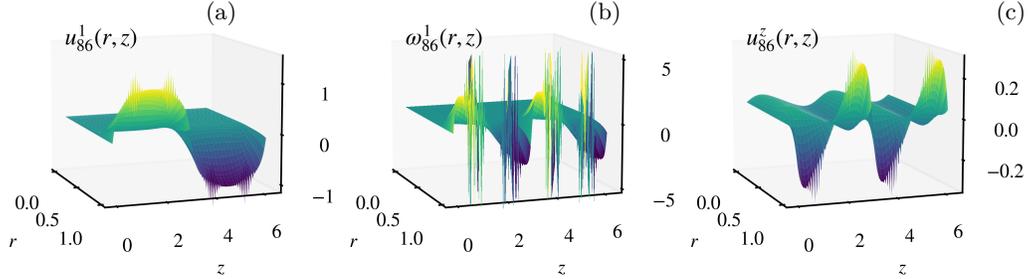

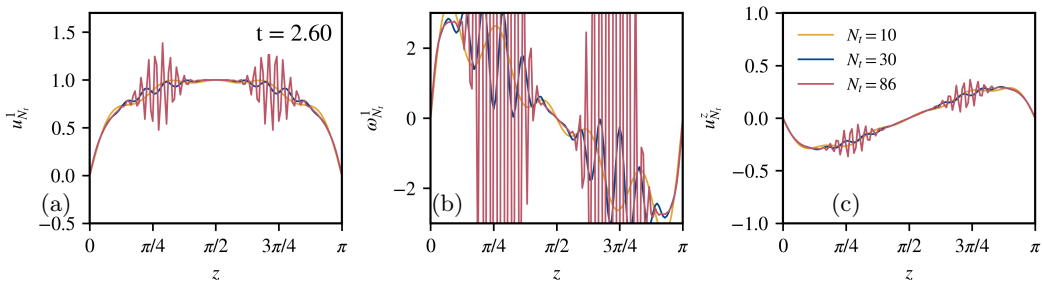
\begin{figure}[htbp]
   \centering
    \begin{tikzpicture}
    \draw (0,0) node[inner sep=0]{\includegraphics[width=\linewidth]{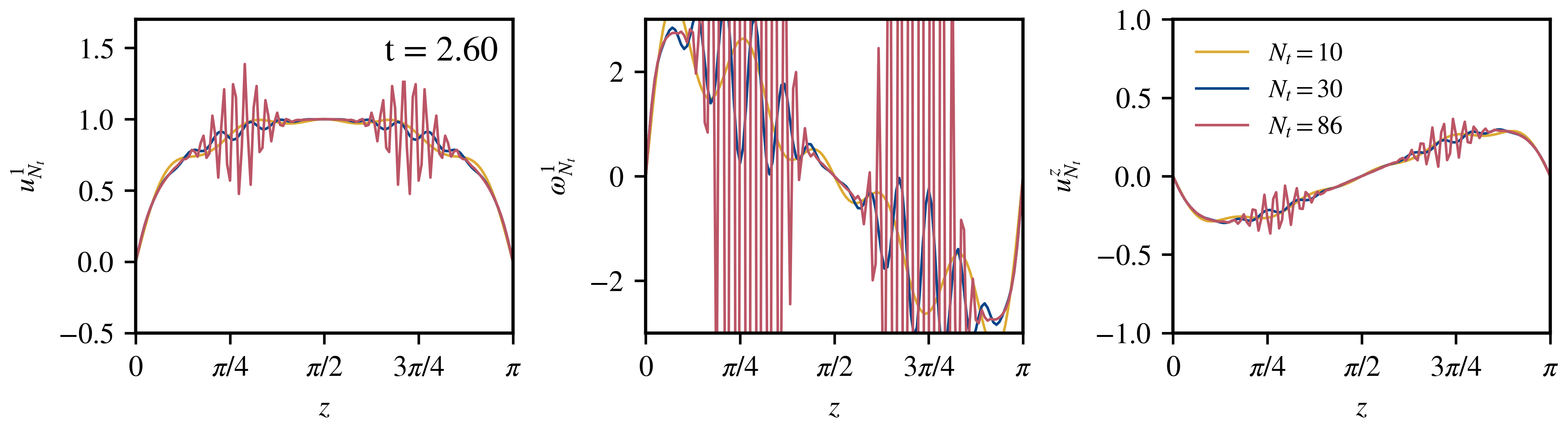}};
    \draw (-6.2,-0.8) node {(a)};
    \draw (-1,-0.8) node {(b)};
    \draw (4.3,-0.8) node {(c)};
    \end{tikzpicture}   
   \caption{Plots vs. $z$ of the truncated time-Taylor series approximations for the fields (a) $u^1_{N_t}(r_N,z,t)$, (b) $\omega^1_{N_t}(r_N,z,t)$, and (c) $u^z_{N_t}(r_N,z,t)$, evaluated at time $t=2.6$ and summed for three orders of truncation, namely, $N_t = 10 \text{ (yellow)}, \ 30$ (blue) and $100$ (red). Early-time resonances occur in all three fields as oscillatory structures localised around $z = \{ \tfrac{\pi}{4}, \tfrac{3\pi}{4} \}$. The oscillations have the largest amplitude for $\omega^1_{N_t}$.}
   \label{fig:3d_Slice}
\end{figure}

In FIG.~\ref{fig:3d_Sum}, we plot, as a function of $(r,z)$, the time-Taylor series truncated at $N_t = 86$ and evaluated at time $t=2.6$ for the fields (a) $u^1_{N_t}(\mathbf{X}_{N,M})$, (b) $\omega^1_{N_t}(\mathbf{X}_{N,M})$, and (c) $u^z_{N_t}(\mathbf{X}_{N,M})$. Early-time resonances emerge at the wall ($r=1$) in all fields. In FIG.~\ref{fig:3d_Slice}, we plot these fields versus $z$, at $r=r_N$. These plots show clearly the development of oscillations at $z=\{ \tfrac{\pi}{4},\tfrac{3\pi}{4} \}$ (in FIG.~\ref{fig:3d_Sum}, also at $\{ \tfrac{5pi}{4},\tfrac{7\pi}{4} \}$). In FIG. 4. of Ref.~\onlinecite{kolluru2022insights} we have reported the emergence of tygers, $z=\{ \tfrac{\pi}{2},\tfrac{3\pi}{2} \}$ in all fields in this model, while using traditional Fourier-Chebyshev pseudospectral methods, for these initial data.
We again emphasize that (as in the 1D HL model) these tygers are different from the early-time resonances in FIG.~\ref{fig:3d_Slice}. 

Given that the even-order coefficients for $u^1_n(r_N,z)$ and the odd-order coefficients for $\omega^1_n(r_N,z), u^z_n(r_N,z)$ are identically zero [FIG.~\ref{fig:3d_coeffs}], the time-Taylor series can be rewritten for the variable $q=t^2$ [cf. Eq.~\eqref{eq:lh1d_t2exp} for the 1D HL model]: 
\begin{subequations}    
\begin{align}
    u^1_{N_t}(r_N,z,q&) = \sum_{n=0}^{N_t/2} u^1_{2n}(r_N,z) {q}^n\,; \\
    \omega^1_{N_t}(r_N,z,q) = q^{1/2} \ \sum_{n=0}^{N_t/2} \omega^1_{2n+1}(r_N,z) {q}^n\,;& \qquad
    u^z_{N_t}(r_N,z,q) = q^{1/2} \ \sum_{n=0}^{N_t/2} u^z_{2n+1}(r_N,z) {q}^n\,.  
\label{eq:t2series_3d}
\end{align}
\end{subequations}

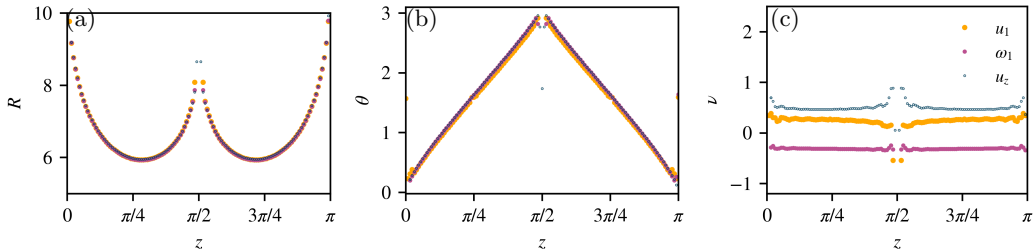
\begin{figure}[htbp]
    \centering
    \begin{tikzpicture}
    \draw (0,0) node[inner sep=0]{\includegraphics[width=0.325\linewidth]{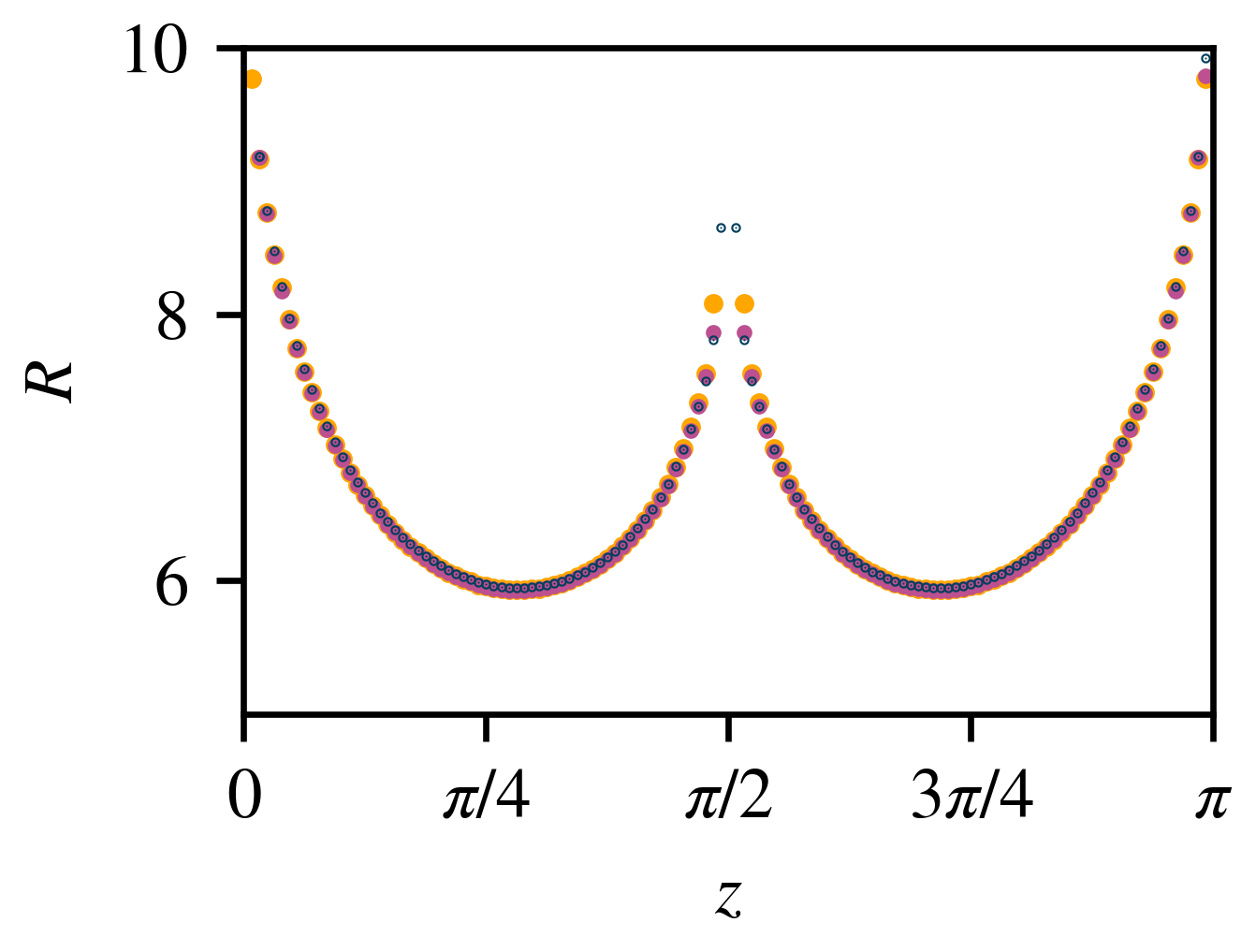}};
    \draw (-1.2,1.45) node {(a)};
    \end{tikzpicture}
    \begin{tikzpicture}
    \draw (0,0) node[inner sep=0]{\includegraphics[width=0.325\linewidth]{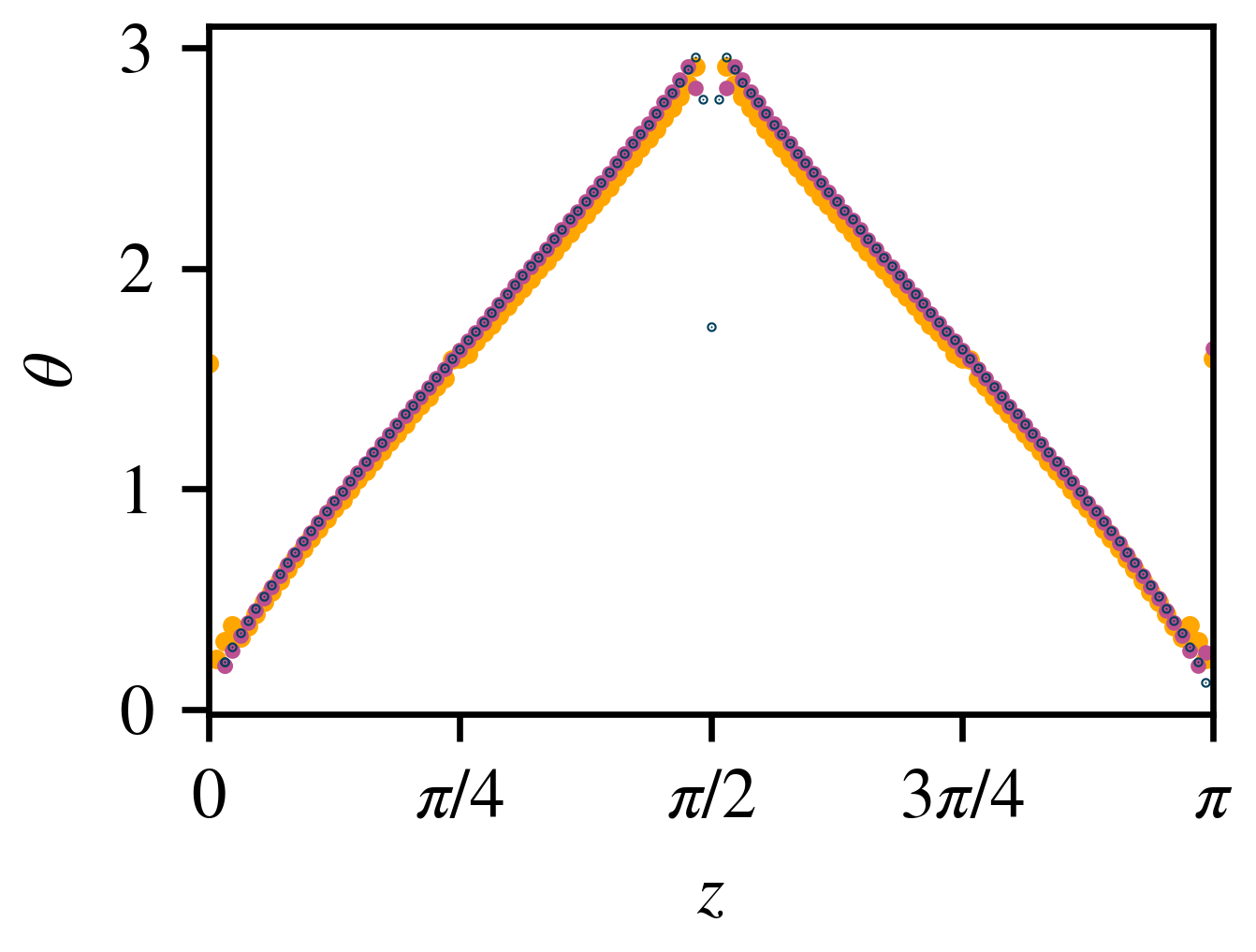}};
    \draw (-1.3,1.45) node {(b)};
    \end{tikzpicture}
    \begin{tikzpicture}
    \draw (0,0) node[inner sep=0]{\includegraphics[width=0.325\linewidth]{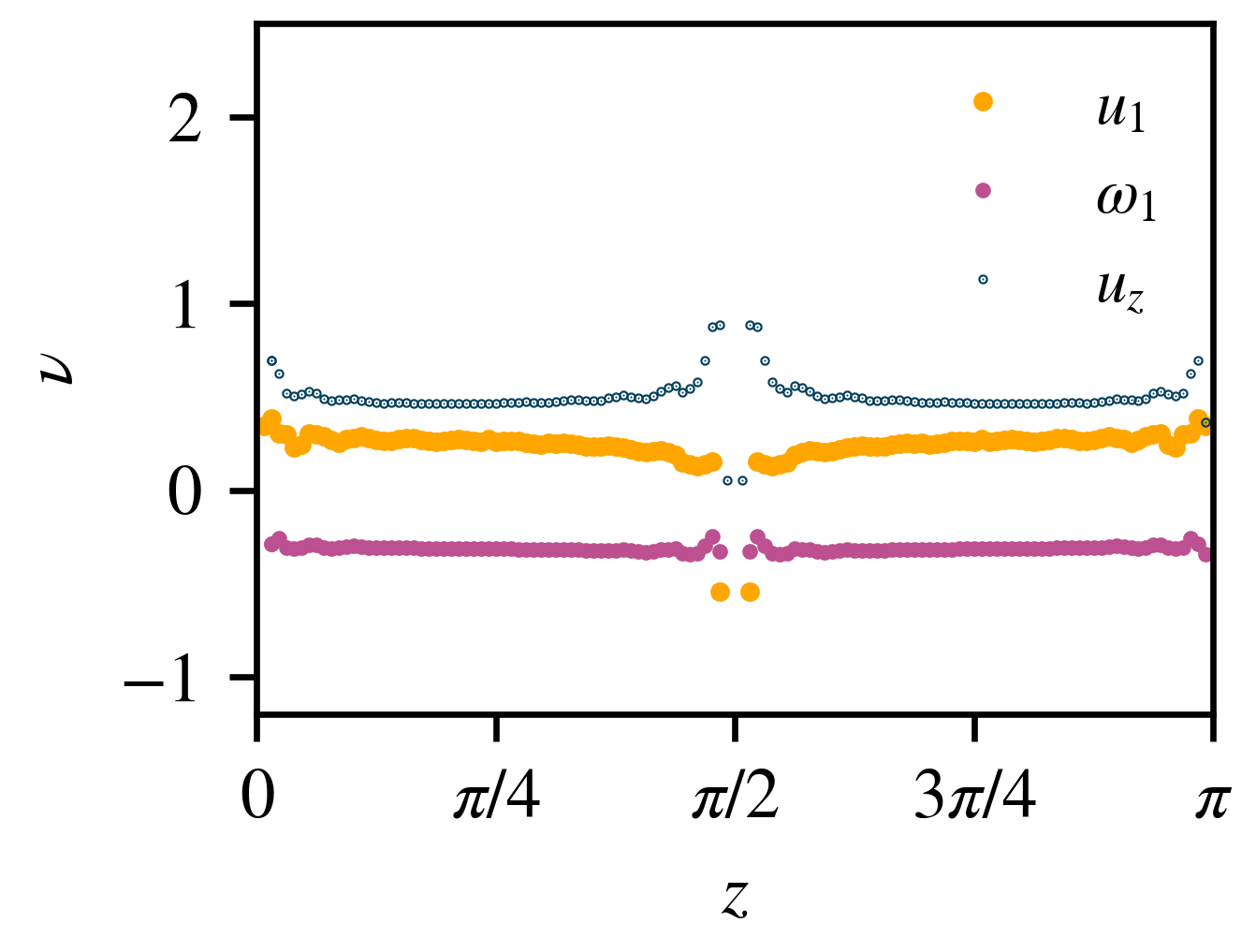}};
    \draw (-1.1,1.45) node {(c)};
    \end{tikzpicture}
    \caption{Plots vs. $z$ of the estimates of (a) $R$, (b) $\theta$, and (c) $\nu$, obtained from $u^1_n(r_N,z) \text{(orange)}, \ \omega^1_n(r_N,z)$(pink) and $u^z_n(r_N,z)$(blue), for $z \in \mathbf{X}_{N,M}$, where $(N=128,M=256)$ and $n\le N_t=86$, using the MR method. A full nonlinear fit of $B_k^2$ in Eq.~\eqref{eq:MR_coeff_text} is performed by using the LMFIT package in Python. For more details on the MR method, see Appendix~\ref{app:MR}.}
    \label{fig:3d_rthetanu} 
\end{figure}

For a given field, we also note that the non-zero time-Taylor coefficients, computed at a point $z$ near the wall $r_N$ by using the initial data~(\eqref{eq:3d_ic}), have a pattern of alternating signs [cf. Sec.~\ref{subsec:Results_1D} for the 1D HL model]. Thus, we can use the MR methods as we did for the 1D HL model in Eq.~\eqref{eq:MR_1d}. For each value of $z \in \mathbf{X}_{N,M}$ near the wall at $r_N$, we construct $B_k^2(z)$ as in Eq.~\eqref{eq:MR_coeff_text}. Our estimates for $R, \ \theta$ and $\nu$, which we obtain by using the LMFIT package~\cite{newville2016lmfit} in Python as for the 1D HL model, are shown in panels (a)-(c) of FIG.~\ref{fig:3d_rthetanu}, where we superpose the estimates for $u^1_{N_t}(r_N,z_j)$ (yellow), $\omega^1_{N_t}(r_N,z_j)$ (pink) and $u^z_{N_t}(r_N,z_j)$ (blue). 

\begin{figure}[htbp]
    \centering    
    \includegraphics[width=0.354\linewidth]{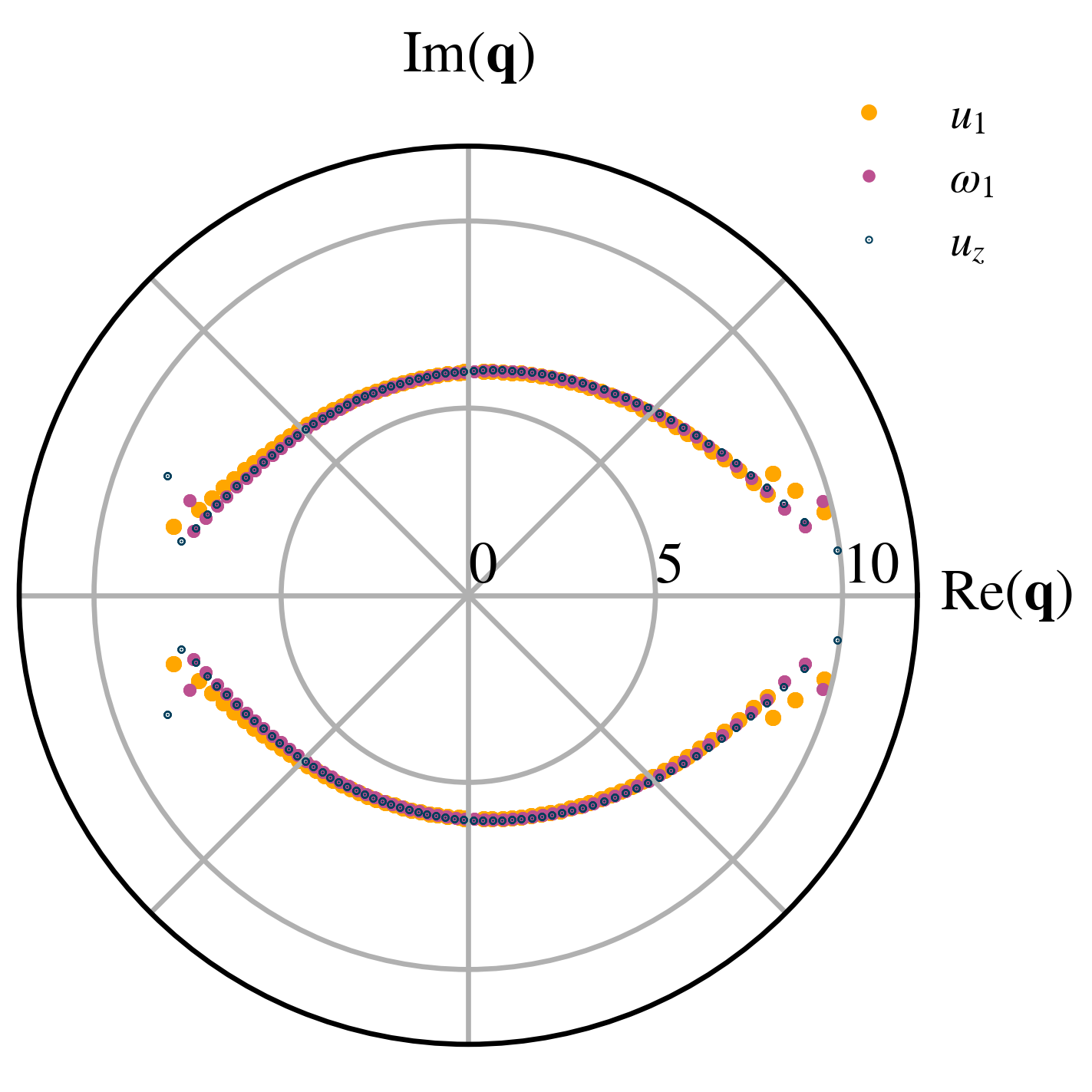}
    \caption{Polar plot in complex-$\mathbf{q}$ plane of $R(z)$ vs. $\theta(z)$ depicting the position of convergence-limiting complex singularities $\mathbf{q}_* = Re^{i \theta}$ as well as their complex-conjugates, for $u^1$ (yellow), $\omega^1$ (pink) and $u^z$ (blue) at $r=r_N$ and $z \in \mathbf{X}_{N,M}$, where $(N=128,M=256)$. The singularities are arranged in the shape of an eye, which is squashed along the $\Im(\mathbf{q})$ axis; the center of the eye is displaced to the right of the origin.}
    \label{fig:3d_eye_q}
\end{figure}

The convergence-limiting singularities for the series, evaluated at the wall for $z = \{ 0, \pi \}$, are located on the positive real axis because $\theta(z)=0$ [cf. FIG.~\ref{fig:1d_eye_q} for the 1D HL model]; these singularities are situated farthest from the origin in the complex-$\mathbf{q}$ plane. The estimates for $R(z)$ in FIG.~\ref{fig:3d_rthetanu}(a) are not arranged symmetrically about $z = \{ \tfrac{\pi}{4}, \tfrac{3\pi}{4} \}$ [unlike in FIG.~\ref{fig:1d_rthetanu}(a) for the 1D HL model]. Here, the convergence-limiting singularities with the smallest radius of convergence [$R_m \simeq 5.95$] occur at $z \simeq 0.88 > \tfrac{\pi}{4}$ and $z \simeq 2.26 < \tfrac{3 \pi}{4}$ in FIG.~\ref{fig:3d_rthetanu} (a). In the complex-$\mathbf{q}$ plane, the arrangement of the convergence-limiting singularities is in the form of an eye [FIG.~\ref{fig:3d_eye_q}], whose center lies on the real axis but is displaced to the right of the origin, for all three fields: $u^1 \text{ (orange)}, \ \omega^1$ (pink) and $u^z$ (blue), because of the asymmetry of $R(z)$ mentioned above. 
{In panel (c) of FIG.~\ref{fig:3d_rthetanu}, we plot the order $\nu$ of the convergence-limiting singularities obtained for the fields $u^1$ (yellow), $\omega^1$ (pink) and $u^z$ (blue). For $u^1$, $\nu$ takes {non-integer} values which depend on $z$; for $z$ near $\tfrac{\pi}{2}$, it is negative but otherwise remains positive. For $\omega^1$, $\nu \simeq -1/3$; and for $u^z$, $\nu \in (0.5,1)$.} [See Appendix~\ref{app:MR} for estimates for $R$ obtained at different distances $r$. ]

\section{Conclusions}
\label{sec:Conclusions}

We have shown how to adapt the methods introduced in Ref.~\onlinecite{rampf} to investigate early-time resonances in two PDEs that have attracted considerable attention recently~\cite{luo2014potentially,luo2014toward,choi2017finite,kolluru2022insights,barkley,hertel} in the context of possible finite-time singularities in ideal hydrodynamical systems. The first is the 1D HL model and the second is the 3DAE; the former is related approximately to the latter [see, e.g., Refs.~\onlinecite{luo2014potentially,choi2017finite,kolluru2022insights}].
We have used initial data from earlier studies of singularities for both these PDEs and employed our pseudospectral Fourier-Chebyshev method~\cite{kolluru2022insights}, with quadruple-precision arithmetic, to compute the time-Taylor series coefficients of the flow fields, up to a high order [$N_t= 100$ (1DHL) and $N_t = 86$ (3DAE)]. The resulting approximations display early-time resonances, which we have studied in detail. In particular, we have demonstrated that the initial spatial location of these resonances is different from that for the tygers, which we have obtained in Ref.~\onlinecite{kolluru2022insights}. We have analysed the time Taylor-series coefficients, by using the MR method, to extract
$R(z)$, $\theta(z)$, and $\nu(z)$, and thence the location and nature of the convergence-limiting singularities. We 
have found that these singularities are distributed around the origin, in the complex-$t^2$ plane, along two curves that resemble an approximately elliptical eye.

The connection between complex-space and complex-time singularity landscapes is a challenging problem that is now being explored for ideal hydrodynamical PDEs such as the inviscid, unforced 1D Burgers equation~\cite{rampf}. Our work has enlarged the scope of such studies by investigating such connections for the 1DHL and 3DAE systems. 
{It is interesting to explore the relationship between the singularity detection for complex-space singularities (e.g, via the analyticity-strip method~\cite{sulem}) and time-Taylor-series methods that we use here. It is important to understand the link (if any) between tygers, in Galerkin-truncated pseudospectral studies, and the early-time resonances, in truncated time-Taylor expansions of solutions of PDEs.} 
The development of numerical schemes for the mitigation of tygers or early-time resonances also poses interesting questions that are being addressed in several recent studies~\cite{murugan2020suppressing,rampf,kolluru2024novel,bessemain}.

\begin{acknowledgments}
We thank SERB, CSIR, and NSM (India) for their support and SERC (IISc) for computational resources.
We thank J.K. Alageshan, U. Frisch, K.V. Kiran, T. Matsumoto, C. Rampf, and S.S. Ray for useful discussions.
\end{acknowledgments}

\bibliographystyle{ieeetr}
\bibliography{references}

\appendix

\section{Code}
\label{app:Code}
 The following \texttt{MATHEMATICA 12} code computes analytical closed form expressions for the time-Taylor coefficients; $u_n(z)$ is stored in \texttt{u[n,z]}, $\omega_n(z) $ in \texttt{w[n,z]} and $v_n(z)$ in \texttt{v[n,z]}. Here, \texttt{Ntmax} represents the order of truncation $N_t$. The initial conditions are given in Lines 4-6. Recursion relations given in Eqs.~\eqref{eq:rec_hl1d} are given in Lines 8-10. The coefficients are then appended to \texttt{listc} which can ultimately be stored symbolically and later retrieved.

\begin{lstlisting}[language=Mathematica,caption={Mathematica 12 code}]
listc={}; Ntmax=70; f[n_] = Floor[(n - 1)/2] ;   
HilbertTransform[f_, x_, X_] :=   Module[ {fp = FourierParameters -> {1, -1}, k},  InverseFourierTransform[ -I (2 HeavisideTheta[k] - 1) *     FourierTransform[f, x, k, fp], k, X, fp]];   
  
u[0, z_] := (Sin[z])^2;
w[0, z_] := 0;
v[0, z_] := Integrate[HilbertTransform[w[0, x], x, z], z];

w[n_, z_] := w[n, z] = If[EvenQ[n], 0,  (- Sum[ If[i != (n - 1 - i), v[i, z] D[w[n - 1 - i, z], z] + v[n - 1 - i, z] D[w[i, z], z], v[i, z] D[w[i, z], z]], {i, 0, f[n], 1}] + D[u[n - 1, z], z])/n ];
u[n_, z_] := u[n, z] = If[OddQ[n], 0,
    (-Sum[ If[i != (n - 1 - i), v[i, z] D[u[n - 1 - i, z], z] + v[n - 1 - i, z] D[u[i, z], z], v[i, z] D[u[i, z], z]], {i, 0, f[n], 1}])/n];
v[n_, z_] :=   v[n, z] = If[EvenQ[n], 0, Integrate[HilbertTransform[w[n, x], x, z], z]];

Do[nn = n;  AppendTo[listc, {n, u[n, z], w[n, z], v[n, z]}], {n, 0,  Ntmax}];
\end{lstlisting}
\section{Mercer-Roberts method}\label{app:MR}

For perturbation series where the coefficients' signs follow a non-trivial pattern, Mercer and Roberts~\cite{mercer1990centre} generalized the Domb-Sykes method to allow for a pair of complex conjugate singularities.

Consider a model function that has complex conjugate singularities at $\mathfrak{t}_*$ and $\overline{\mathfrak{t}_*}$
\begin{equation}
	\mathfrak{u}(\mathfrak{t}) = \left( 1- \frac{\mathfrak{t}}{\mathfrak{t_*}} \right)^{\nu} +\left( 1- \frac{\mathfrak{t}}{\overline{\mathfrak{t_*}}} \right)^{\nu}; \quad \text{where } \mathfrak{t_*} \coloneqq R e^{i \theta}   
	\label{eq:modelfn_defn}
\end{equation}

 For $|\mathfrak{t}|<R$, the model function $\mathfrak{u}$ has the following Taylor expansion around $t=0$:
\begin{equation}
	\mathfrak{u}(\mathfrak{t}) = \sum_{n=0}^{\infty} 2(-1)^n \binom{\nu}{n} R^{-n} \cos(n \theta) \mathfrak{t}^n
	\label{eq:modfn_ts}
\end{equation}

\begin{figure}[htbp]
    \centering
     \begin{tikzpicture}
    \draw (0,0) node[inner sep=0]{\includegraphics[width=0.45\linewidth]{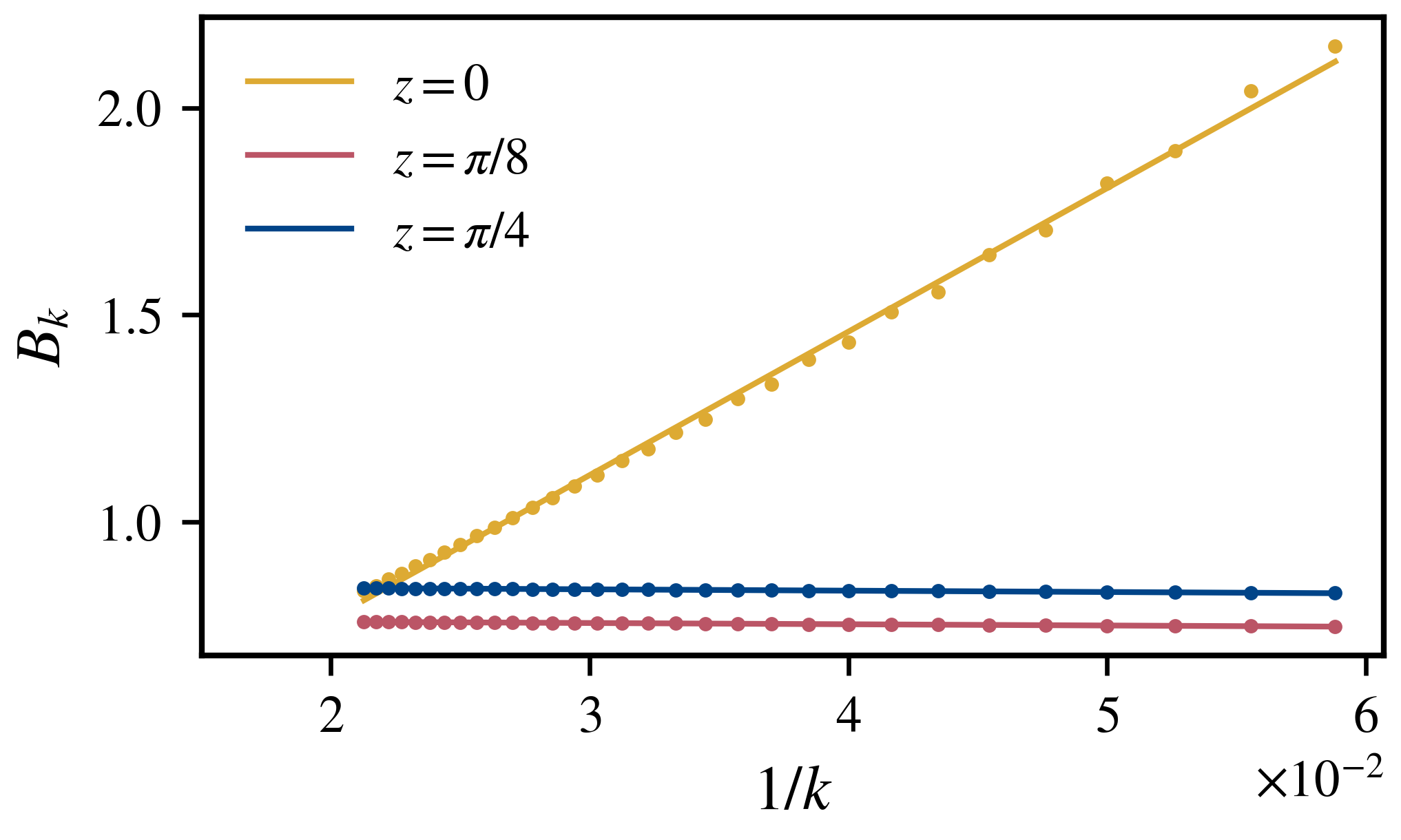}};
    \draw (-2.2,2.5) node {\large{(a)}};
    \end{tikzpicture}
    \hspace{1cm}
         \begin{tikzpicture}
    \draw (0,0) node[inner sep=0]{\includegraphics[width=0.4\linewidth]{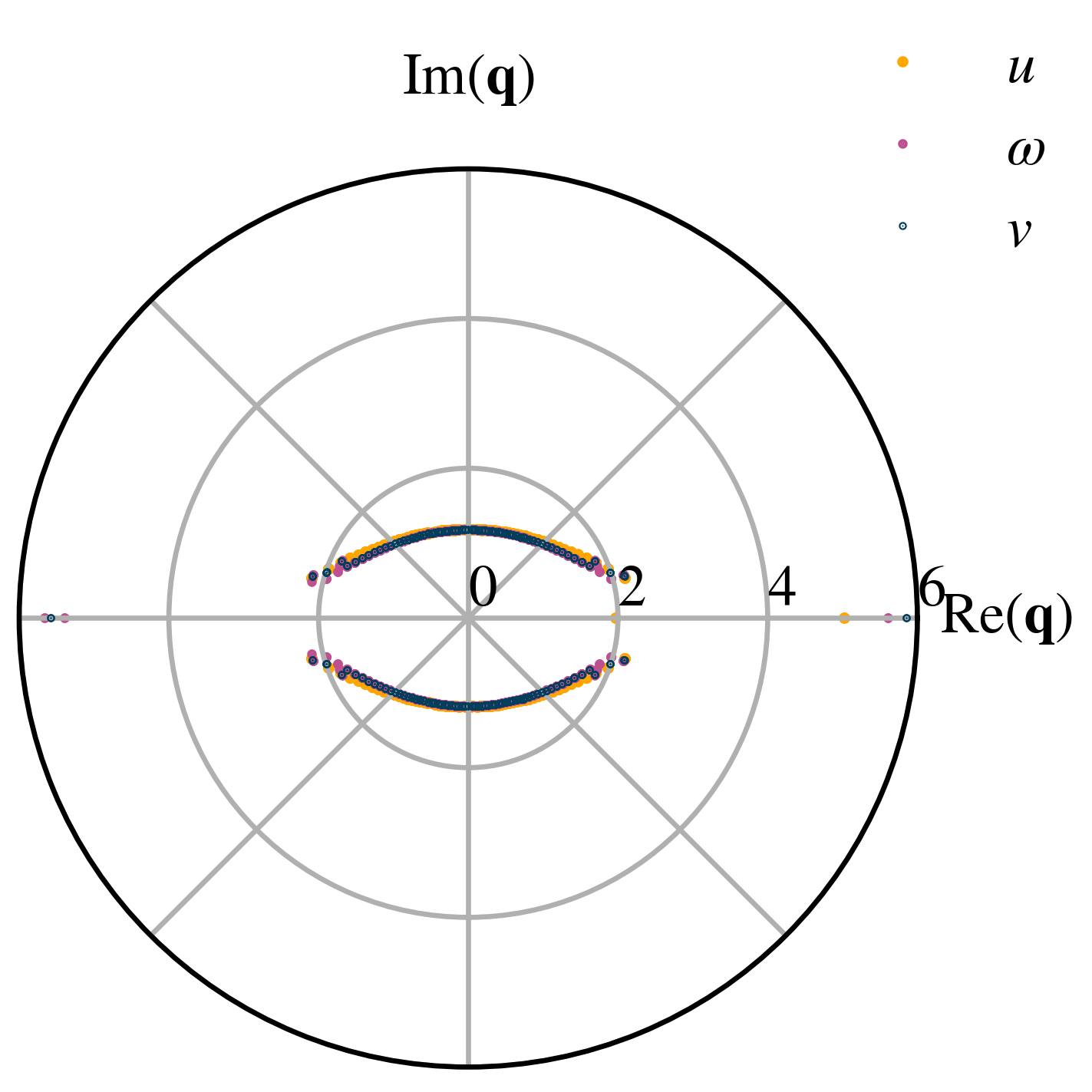}};
    \draw (-2.5,2) node {\large{(b)}};
    \end{tikzpicture}
    \caption{\textbf{For the 1D HL model:} (a) Plots vs. $1/k$ of $B_k(z)$ computed from $u_n(z)$ using Eq.~\eqref{eq:MR_coeff_text}, for $z=0 \text{ (yellow)},\pi/8\text{ (red)},\pi/4\text{ (blue)}$. 
    (b) Extended polar plot in complex-$\mathbf{q}$ plane of $R(z)$ vs. $\theta(z)$ depicting the position of convergence-limiting complex singularities $\mathbf{q}_* = Re^{i \theta}$ as well as their complex-conjugates, for $u$ (yellow), $\omega$ (pink) and $v$ (blue). Here, we note the singularities on the real-axis obtained using MR method for this model.    
    The time-Taylor coefficients are estimated on a uniform grid of $\mathbf{X}_{N=256}$ using quadruple-precision Fourier pseudospectral methods.}
    \label{fig:lh1d_app}
\end{figure}

Mercer and Roberts showed that $R, \theta$ and $\nu$ can be determined by relating the model function in Eq.~\eqref{eq:modfn_ts} to the original series $u(t) = \sum_{n=0}^{\infty} u_n t^n$. For each $4$-tuple of $u_n$, we construct:
\begin{subequations}
  \label{eq:MR_coeff}
	\begin{align}
		B^2_k &= \frac{u_{k+1} u_{k-1} - u^2_k}{u_k u_{k-2} - u^2_{k-1} } \\
		cos \theta_k &= \frac{1}{2} \left( \frac{u_{k-1}B_k}{u_k} + \frac{u_{k+1}}{u_k B_k}  \right)\,,
	\end{align}
\end{subequations}
where $k= 2, 3, \ldots$. The leading-order behaviour is then obtained by substituting Eq. ~\eqref{eq:modfn_ts} into Eq.~\eqref{eq:MR_coeff}:
\begin{subequations}  \label{eq:modcoeffs_l}
	\begin{align}
		B_k =& \frac{1}{R} - \left( \frac{\nu +1}{R} \right) \frac{1}{k} + \left( \frac{\nu+1}{2R} \frac{ \sin (2k-1) \theta }{\sin \theta} \right) \frac{1}{k^2} + \mathcal{O}(\frac{1}{k^3}) \\
		\cos \theta_k &= \cos \theta + \left[ \cos \theta \ (\nu +1) \ \left( 1 - \frac{\cos (2k-1)\theta}{\cos \theta} \right) \right] \frac{1}{k^2} + \mathcal{O}(\frac{1}{k^3})\,.
	\end{align}
\end{subequations}
For large $k$, we can extract $1/R$ from the intercept on the vertical axis,in the plot of $B_k$ vs. $1/k$, and $\cos \theta$ from the plot of $\cos \theta_k$ vs. $\frac{1}{k^2}$. In FIG.~\ref{fig:lh1d_app}(a), we show the plots versus $1/k$ of $B_k$, obtained from $u_n(z)$ in Eqs.~\eqref{eq:rec_u} for the 1D HL model, at the spatial locations $z=\{ 0,\tfrac{\pi}{8}, \tfrac{\pi}{4} \}$ in yellow, red, and blue, respectively. For these values of $z$, the leading-order linear behaviour dominates asymptotically. For $\theta$ close to $\{0,\pm \pi\}$, the estimators in Eq.~\eqref{eq:modcoeffs_l} become unreliable as high-orders terms are not negligible. 

{In FIG.~\ref{fig:lh1d_app}(b), we show the eye, including the MR estimates obtained by using the linear forms (FIG.~\ref{fig:lh1d_app}(a)), exactly at the singular $z$ points.}

\begin{figure}
    \centering
         \begin{tikzpicture}
    \draw (0,0) node[inner sep=0]{\includegraphics[width=0.45\linewidth]{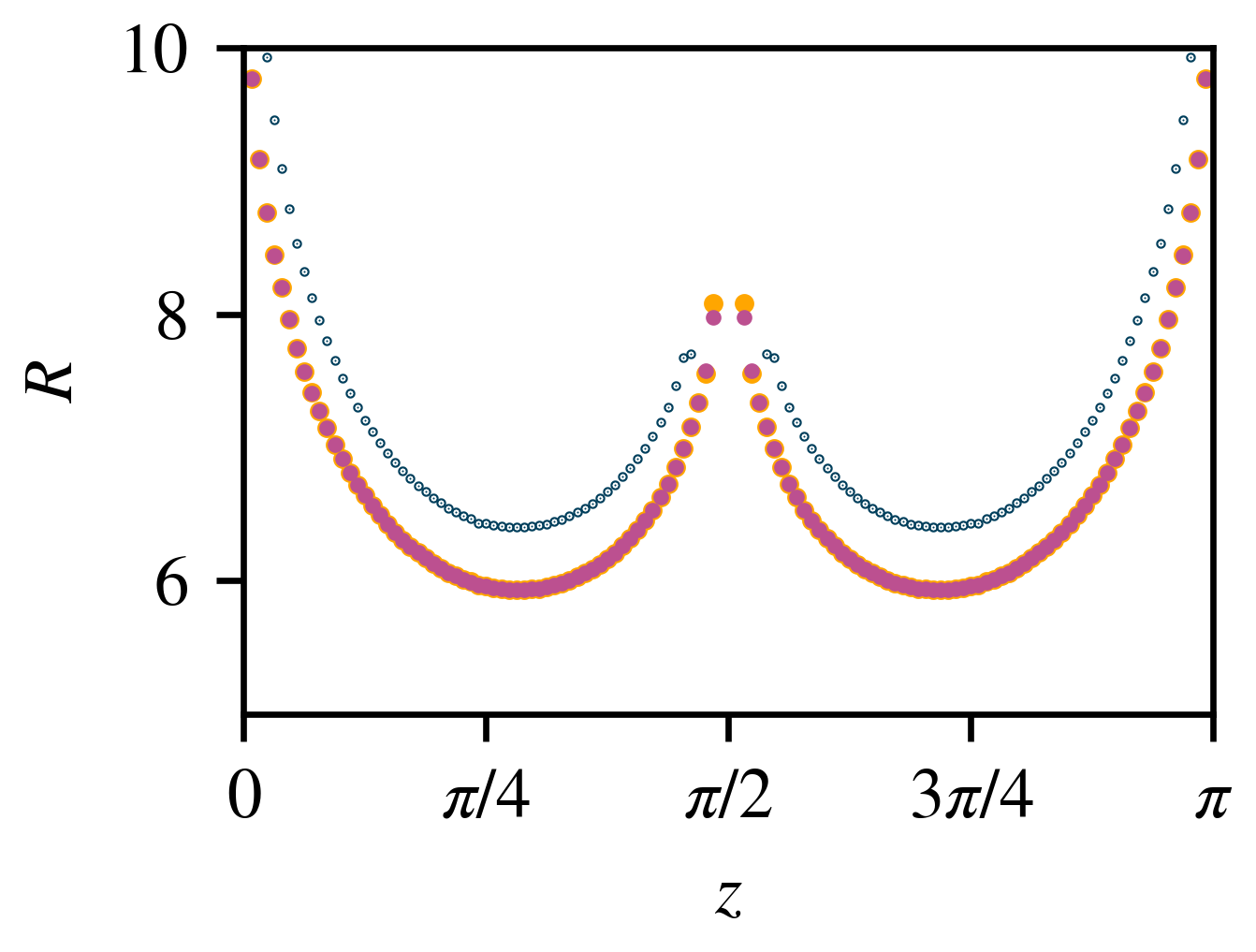}};
    \draw (-2.1,3) node {\large{(a)}};
    \end{tikzpicture}
    \hspace{1cm}
         \begin{tikzpicture}
    \draw (0,0) node[inner sep=0]{ \includegraphics[width=0.45\linewidth]{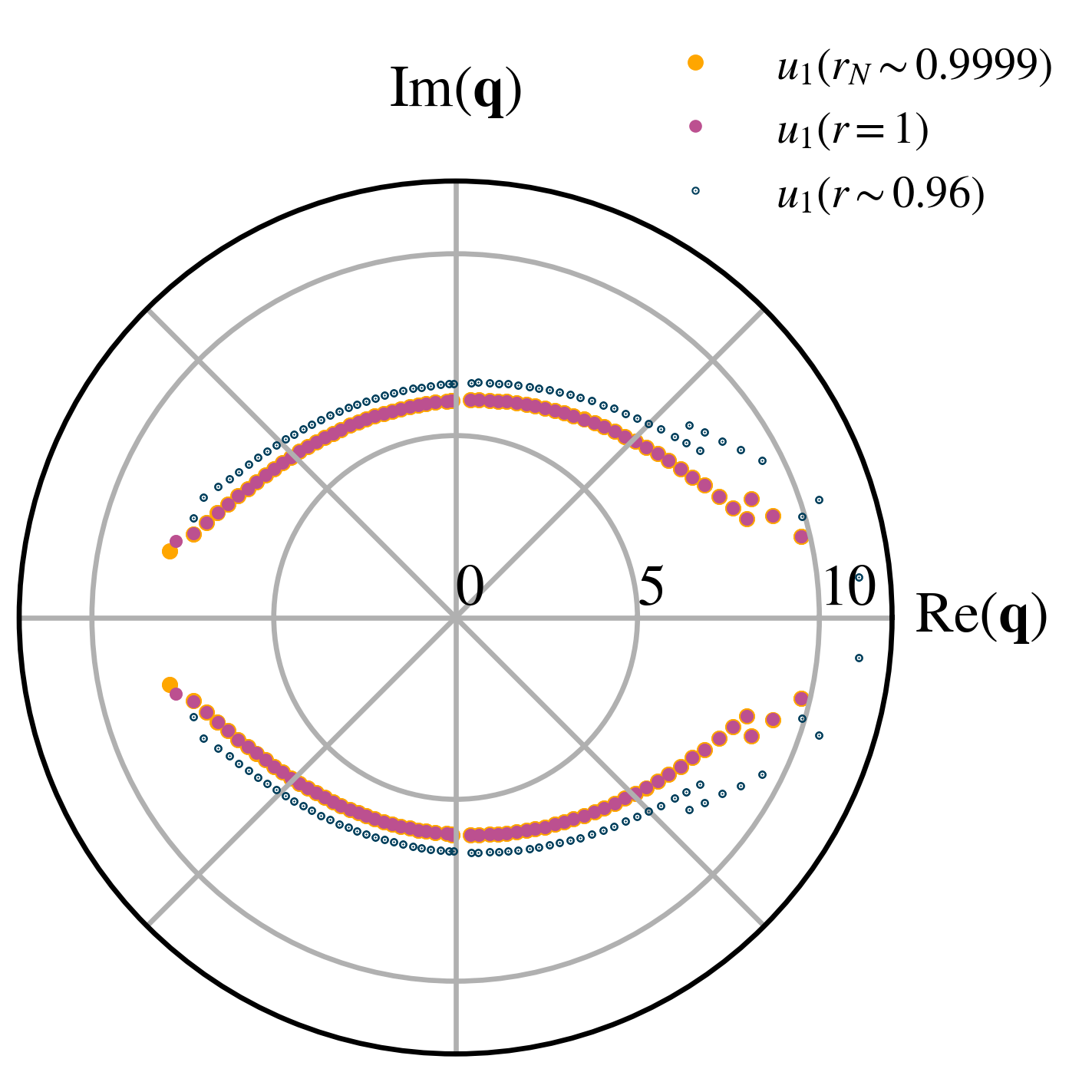}};
    \draw (-2.5,2.6) node {\large{(b)}};
    \end{tikzpicture}
    \caption{\textbf{For 3DAE:} (a) Plots vs. $z$ of the estimates of $R$ and, (b) polar plot, in the complex-$\mathbf{q}$ plane, of $R(z)$ vs. $\theta(z)$ depicting the position of convergence-limiting complex singularities $\mathbf{q}_* = Re^{i \theta}$ as well as their complex-conjugates, obtained from $u^1_n(r_N,z)$ for $z \in \mathbf{X}_{N,M}$, where $(N=128,M=256)$ and $n\le N_t=86$, using the MR method; the estimates are shown for different distances from the wall at $r \in \{ 0.96, \, r_N , \, 1 \}$ (see legend). Here, we see that the eye is shifted for all $r$ and the MR estimates worsen far from the wall (where the singularity precipitates). }
    \label{fig:lh3d_app}
\end{figure}

{In FIG.~\ref{fig:lh3d_app} (a), we show the estimates for $R$, obtained as above, at different distances $r$ from the wall. In particular, we compute the time-Taylor approximations at the wall $r=1$, at $r=r_N$ (as before), and at an interior point $r=0.96$. All these values of $r$ yield eyes with shifted centers of $R(z)$ about $z = \{ \tfrac{\pi}{4}, \tfrac{3\pi}{4} \}$ persists. In FIG.~\ref{fig:lh3d_app} (b), we plot the associated eyes in the complex-$\mathbf{q}$ plane; the center of the eye is shifted from the origin for all the choices of $r$ above. To the best of our knowledge, the eye associated with the 3DAE is the first with a shifted center; this shift remains to be understood.}

\end{document}